\documentclass[12pt]{article}
\usepackage{geometry}
\usepackage{a4}
\usepackage{graphicx,subfigure}
\usepackage{epsf}
\usepackage{amsmath}
\usepackage{amssymb}
\usepackage{cite}
\usepackage{tensor}
\begin{document}
\def\rh{r_H}
\def\rt{r_t}
\def\onefour{\frac{1}{4}}
\def\half{\frac{1}{2}}
\def\threefour{\frac{3}{4}}
\def\fivefour{\frac{5}{4}}
\def\sevenfour{\frac{7}{4}}
\def\mfour{\frac{m}{4}}
\def\nfour{\frac{n}{4}}
\def\s{{\mathbb{S}}}
\def\T{{\mathbb{T}}}
\def\Z{{\mathbb{Z}}}
\def\W{{\mathbb{W}}}
\def\Bbb{\mathbb}
\def\BZ{\Bbb Z} \def\BR{\Bbb R}
\def\BW{\Bbb W}
\def\BM{\Bbb M}
\def\BC{\Bbb C} \def\BP{\Bbb P}
\def\CP{\BC\BP}
\begin{titlepage}
\title{Distinguishing between Kerr and rotating JNW space-times via frame dragging and tidal effects}
\author{}
\date{
Tathagata Karmakar, Tapobrata Sarkar
\thanks{\noindent E-mail:~ karmakar, tapo@iitk.ac.in}
\vskip0.4cm
{\sl Department of Physics, \\
Indian Institute of Technology,\\
Kanpur 208016, \\
India}}
\maketitle\abstract{
\noindent
In this paper, we first investigate some aspects of frame dragging in strong gravity. The computations
are carried out for the Kerr black hole and for the rotating Janis-Newman-Winicour solution, that is known to have a naked singularity
on a surface at a finite radius. For the Kerr metric, a few
interesting possibilities of gyroscope precession frequency, as measured by a Copernican observer outside the ergoregion, are pointed out. 
It is shown that for certain angular velocities of a stationary observer, this frequency might vanish exactly,
close to the ergoregion. Similar 
computations are repeated for static observers in the naked singularity background, and strong enhancement of the 
Lense-Thirring precession frequency compared to the black hole case is established. Then, 
we study the nature of tidal forces in the rotating naked singularity background, in Fermi normal coordinates. 
Here, physical quantities characterizing 
tidal disruptions of celestial objects in equatorial circular orbits are computed numerically. Our results here
indicate that there might be significant deviations from 
corresponding Kerr black hole calculations, up to the level of approximation that we consider.}
\end{titlepage}
\section{Introduction}

Singular solutions of Einstein's General Relativity (GR) \cite{Weinberg},\cite{Chandra},\cite{Wald},\cite{Hartle},\cite{Poisson} have been 
at the focus of attention over the last
few decades. These include black holes, with the singularity hidden by an event horizon, or naked singularities, objects that do not have
a horizon to cover the singularity. The relevance of these objects are immediate as it is well known that at the center of galaxies 
lie very massive compact objects which are believed to be black holes. The cosmic censorship conjecture, proposed by Penrose
states that nature abhors a singularity, i.e naked singularities are not allowed by a physical collapse process. This is however yet to
be proved completely, and indeed, there has been substantial interest in gravitational collapse processes that might give rise to
a singularity without an event horizon. 

Several interesting aspects have been revealed by such analyses. As was shown by Joshi and collaborators (see, e.g \cite{Joshi1},\cite{JoshiBook}),
the end stage of gravitational collapse might be driven towards a naked singularity in the presence of non-zero pressures. Various properties
of such equilibrium stable solutions have been investigated in the recent past. Recently, it has been shown \cite{Tapo1} that naked singularities
as the end stage of spherical gravitational collapse is also possible in a cosmological scenario. The issue of naked singularities assume importance
in the context of quantum theories of gravity, as singularities in GR indicates its breakdown, and points to its limits of applicability. 
Whether quantum effects set in at the late stage of gravitational collapse and modify such limits is little understood, and promises to be 
of great relevance in the future. 

Several works have appeared in the literature in the recent past, that try to distinguish between observational properties of black holes and 
naked singularities, for example via their accretion disc properties, gravitational lensing, etc (see, e.g \cite{Joshi2}, \cite{Gyulchev},\cite{Harko}). 
The purpose of this paper is to elaborate upon and extend such analyses. In particular, we consider rotating solutions of Einstein's equations
and focus on two effects over here -- the Lense-Thirring (L-T) effect and the tidal effect. The former is a profound ``frame dragging'' effect due
to a stationary space-time, and the latter gives an indication of the (in)stability of rotating celestial bodies near a singularity. In particular,
we will consider the strong-field Lense-Thirring effect \cite{Straumann},
recently studied for the Kerr black hole \cite{ParthaDa},\cite{JoshiLT1},\cite{JoshiLT2} (see also \cite{Chakraborty1}, \cite{Chakraborty2}
for related works). 
In the first part of this paper, we study this object more closely and point out some interesting physical features that might arise close to (but outside)
the ergoregion of Kerr black holes. Two important observations in this context are that firstly, the L-T precession frequency can 
speed up as the black hole slows down (due to a Penrose process) and secondly, for stationary observers on the equatorial plane, the 
precession frequency
of a test gyroscope vanishes (without approximation) outside the ergoregion for some values of the observer's angular velocity. 
The study is then repeated for the rotating Janis-Newman-Winicour (JNW) solution of 
GR \cite{KroriBhattacharjee} previously considered in \cite{Gyulchev},\cite{Harko} 
and a comparitive analysis is performed with the black hole case. We show that under certain restrictive assumptions, the Lense-Thirring
effect might be an indicator of differences between black holes and naked singularity backgrounds. 

In the second part of this paper, we study the tidal (i.e non-local) forces on celestial objects, specialized to the rotating JNW backgrounds. 
This is done in Fermi normal coordinates \cite{Manasse},\cite{Poisson} which are locally flat coordinates along a geodesic. In the context of 
Kerr black holes, this computation was performed by Ishii \cite{Ishii}, building upon the previous work of \cite{Fishbone}. In \cite{Ishii}, 
the computation was done up to fourth order in Fermi normal coordinates, by taking into account the internal fluid dynamics of the celestial 
object which is acted upon by the black hole background. Here, in an effort to again contrast physical phenomena in black holes and 
naked singularities, we perform such an analysis with the background being the rotating JNW solution. 

We now summarize the layout and main findings of this paper. After a (very) brief review of the relevant concepts in the next section, in 
section 3, we first consider strong field L-T precession in Kerr and rotating JNW backgrounds. The former has been considered 
recently (see, e.g \cite{ParthaDa}, \cite{JoshiLT2}) mostly in the context of ``Copernican'' observers (to be explained later) 
inside the ergoregion of the black hole. We show here that for a static 
observer, there might be interesting physics outside the ergoregion as well, and that as the black hole's angular 
momentum is lowered, L-T precession might
speed up, in direct contrast with a weak field L-T analysis \cite{Hartle}. The lowering of the black hole angular momentum might arise due to 
a Penrose process (see, e.g \cite{Townsend}) and we will see that the speeding up of the L-T frequency might be possible in this physical
situation. We contrast this with the accreting Kerr black hole using a result due to Bardeen \cite{BardeenKerr} and show that this 
is an impossibility. We then consider a generic stationary observer outside the ergoregion of the Kerr black hole, and show that
(under certain realistic assumptions), the precession frequency of a test gyroscope might vanish exactly, as the angular momentum of the 
background changes. This phenomenon persists for an accreting black hole. 

Section 4 deals with our other topic of interest, namely tidal forces. We compute the tidal disruption limits in rotating JNW backgrounds
for equatorial circular orbits. This is done numerically in Fermi normal coordinates. Finding such coordinates for a stationary space-time in general 
involves a lengthy calculation. Here, we first offer a simple and easy prescription for achieving this on the equatorial
plane of a sufficiently generic stationary space-time. Applying this to the rotating JNW background, after giving the details of the numerical procedure
used, it is seen that when we are far from the Kerr limit, 
the tidal forces might show some peculiar behavior in rotating naked singularity backgrounds compared to their black hole cousins. 
While indicative of new physical phenomena, we point out that this might be a limitation of the order of approximation used, and comment
on the results. Finally, section 5 closes this paper with a discussion of the main findings here. 

\section{Basic Notations and Conventions}

The purpose of this section is to motivate the topics that will be of our interest in this paper. We will be brief here, as the 
material here is mainly intended to serve as a short review, and more
details can be found in references contained herein. We start with the L-T precession in stationary backgrounds. 

\subsection{Lense-Thirring precession in stationary backgrounds}

The Lense-Thirring precession is a profound effect of rotating space-times on a test gyroscope. 
A rotating space-time drags a locally inertial frame, causing the spin of the gyroscope to precess. In a weak 
field limit, the computation of this rate is a textbook exercise, and gives (eq.(14.34) of \cite{Hartle}), with ${\vec J}$
being the angular momentum of the rotating space-time :
\begin{equation}
{\vec \Omega_{LT}} = \frac{1}{r^3}\left(3({\vec J}\cdot{\hat r}){\hat r} - {\vec J}\right)
\label{weakfieldLT}
\end{equation}
When ${\hat r}$ points along ${\vec J}$, this reduces to the expression $\Omega_{LT} = 2J/r^3$. This is the known
inverse-cubed behaviour of the L-T frequence with distance. In particular, when $J$ increases or decreases (with the
weak-field limit still assumed to be valid), the linear increase or decrease of $\Omega_{LT}$ is evident. We note here
that a second effect, called the geodetic effect or the de-Sitter precession of a test gyroscope is very much relevant in
this context, and we will comment on this shortly in the context of stationary observers. 

The weak field computation is of course of limited interest in strong gravity regimes which is our main 
interest here, and there one has to use a more general
expression. In fact, an exact expression for the L-T precession frequency has been derived in the textbook by Straumann \cite{Straumann}.
This is done in a special frame at rest with respect to fixed stars, called a Copernican frame. The result for the precession
vector in this frame is (eq.(1.161) of \cite{Straumann}) : 
\begin{equation}
\Omega_{LT} = \frac{g_{00}}{2\sqrt{-{\rm det}~g}}\epsilon_{ijk}\left(\frac{g_{0i}}{g_{00}}\right)_{,~j}
\left(g_{lk} - \frac{g_{l0}g_{k0}}{g_{00}}\right)dx^k
\label{LTMain}
\end{equation}
where a comma denotes a differentiation and ${\rm det}~g$ is the determinant of the metric. 
This will be our main formula for computation of the L-T frequencies in what follows. We remind the reader
that this formula refers specifically to the Copernican frame, as will be the case here. One can generically consider
a static or a stationary observer in the background of a rotating space-time. The four velocity of such observers
are given by \cite{Hartle}
\begin{equation}
u = u^t\left(1,0,0,0\right),~~~u = u^t\left(1,0,0, \Omega\right)
\label{fourvels}
\end{equation}
respectively, where the condition $u^{\mu}u_{\mu}=-1$ provides restrictions on the possible values of the observer's
angular velocity $\Omega$. We point out that for stationary Copernican observers,
the computed frequency in the frame of the observer is the total frequency of a test gyroscope, which includes
the L-T precession, the de-Sitter effect, etc. 
The corresponding formula for the precession frequency is more elaborate than that of eq.(\ref{LTMain}) to which it
reduces in the limit $\Omega \to 0$ (see eq.(14) of \cite{JoshiLT2}). 
It is also well known that static observers cannot exist inside the ergoregion of the Kerr black hole. We will however
be mostly interested in observers outside the ergoregion, for which this situation does not arise. 

The importance of the L-T precession cannot be over emphasized. We simply remind the reader that the L-T
precession frequency (along with the geodetic precession) were recently measured by the gravity probe B
experiment \cite{GPB} and confirmation with theoretical predictions is hailed as a further success of Einstein's GR. 
A discussion of the L-T effect in black holes and naked singularity backgrounds will be the subject of the section 3. 

\subsection{Tidal Forces in stationary backgrounds}

We now briefly review the idea of tidal forces, which arise out of non-local gravitational interactions. 
Let us consider, a celestial object (maybe a star) moving under the effect of gravity of a central massive object (which might 
exemplify a singularity). Non local gravitational effects might cause this object to break apart. In the Newtonian context, this is not difficult
to see, and is indeed a textbook exercise.
Let us take a massive central object of mass $M$ and radius $R$ and a star (consisting of incompressible matter, say), 
of mass small $M_s$ and small radius $r_s$. A reasonable approximation that considerably simplifies 
calculations is to assume that both objects are
spherical, and their centres are say at a distance $d$ apart.  
If the star is small compared to the central object, so that $r_s/R \ll 1$ and equating the gravitational ``stretching force''
with the self gravity of the small star shows that 
the latter will break apart once the (dimensionless) ratio  $(M_s/M)(d/R_s)^3$ goes below a critical value. 
This value will, in general, depend on the spin angular velocity of 
the central object as well as the star. This ratio given above translates into the fact that a star gets tidal disrupted (or breaks apart) 
once the radial distance between the star and the central massive object goes below some
critical value. This critical distance is popularly known as the Roche limit. 

It is in general substantially more difficult to perform the same analysis in the context of GR, which will be our focus in this paper.
This is due to the subtle interplay of the fluid dynamics of the star interior with the tidal force. Further, one needs to perform the computations
in the Fermi normal coordinate system which is a locally inertial system along a geodesic (being traced out by the star) 
and this itself becomes complicated for stationary backgrounds. Fortunately, both problems can be addressed, if one restricts to
an equatorial plane. In this plane, using up to fourth order in Fermi normal coordinates, the tidal forces in a Kerr background was 
computed in \cite{Ishii}. Here, the star is treated in a Newtonian approximation. 

In \cite{Tapo2}, a comparison was made between the nature of tidal forces in static naked singularity and black hole backgrounds.
There, a few known static naked singularity backgrounds were considered and it was shown that these might have substantial effects
on tidal forces, as compared to black holes. For example, it might be possible for a neutron star to disintegrate in a naked singularity
background at a certain radial distance, while it might be stable in a black hole background at the same radius. 
In this paper, we compute tidal effects in the rotating JNW background. As a byproduct, we also give a relatively easy method to compute
Fermi normal coordinates for a generic stationary space-time. 

We now proceed to the main part of this paper. 

\section{Strong field Lense-Thirring precession in Kerr and rotating JNW backgrounds}

We will consider the Kerr metric written in Boyer-Lindquist coordinates,\footnote{We will, in general, work in geometrized units where the 
Newton's constant $G$ and the speed of light $c$ are both set to unity. We will comment upon the restoration of units in the next
section.}
\begin{eqnarray}
ds^2 &=& - \left(1 - \frac{2Mr}{\gamma \rho^2}\right) \left(dt-\omega d\phi \right)^2 +
\rho^2\left(\frac{dr^2}{\Delta}+d\theta^2+\sin^2\theta d\phi^2\right)\nonumber\\
&-& 2w(dt-\omega d\phi)d\phi
\label{Kerrmetric}
\end{eqnarray}
where $M$ is the Arnowit-Deser-Misner (ADM) mass, and we have defined
\begin{equation}
\omega=a\sin^2\theta~,~\rho^2=r^2+a^2\cos^2\theta~,~\Delta=r^2+a^2-2Mr~.
\label{eq2a}
\end{equation}
Note that $a = J/M$ is a convenient rotation parameter with $J$ being the angular momentum of the black hole 
(we will sometimes loosely refer to $a$ as the angular momentum but the meaning should be obvious from the context). 
The Kerr black hole is extremely well studied and we will simply list the relevant quantities that will be useful for our analysis
to follow (see, e.g \cite{Visser}). These are
\begin{eqnarray}
&& r_{\pm} = M \pm \sqrt{M^2 - a^2},~~~{\rm Outer~and~inner~horizons}\nonumber\\
&& r^E_{\pm} = M \pm \sqrt{M^2 - a^2\cos^2\theta},~~~{\rm Outer~and~inner~ergosurfaces}~.
\label{conds}
\end{eqnarray}
In addition, there is a ring singularity which, in cartesian coordinates, is given by $z=0,~x^2+y^2=a^2$. The ergoregion, defined as
the region between the outer ergosurface $r^E_+$ and the outer event horizon $r_+$ is a special region of space-time where it is not
possible to define a static time-like observer.
Also, from Eq.(\ref{conds}), it follows that in order to avoid a naked singularity, we should demand that $|a| \leq M$, with the equality holding 
for an extremal Kerr black hole. 

The Lense-Thirring precession frequency is known in the literature (see, e.g \cite{ParthaDa},\cite{JoshiLT2}) and from eq.(\ref{LTMain})
its magnitude can be readily calculated to be 
\begin{equation}
\Omega_{LT} = \frac{a M \sqrt{\mathcal A}}{\left(a^2 \cos ^2(\theta )+r^2\right)^{\frac{3}{2}} \left(a^2 \cos ^2(\theta )+r
(r-2 M)\right)}~.
\label{LTKerr}
\end{equation}
Here, ${\mathcal A}$ is defined as  
\begin{equation}
{\mathcal A} = a^4 \sin ^2(\theta ) \cos ^4(\theta )+r^2 \cos ^2(\theta )
\left(a^2 \cos (2 \theta )+3 a^2+4 r (r-2 M)\right)+r^4 \sin ^2(\theta)
\end{equation}
In the weak field limit, i.e for $r \gg M$, the above equation reduces to 
\begin{equation}
\Omega_{LT} = \frac{aM}{r^3}\sqrt{\left(4\cos^2\theta + \sin^2\theta\right)} + {\mathcal O}\left(\frac{1}{r^4}\right).
\label{weak}
\end{equation}
This is as expected : increasing the rotation parameter linearly increases the Lense-Thirring precession frequency, which falls off as 
an inverse cube of the distance. 
Important properties of the LT precession of test gyrocsopes have been considered in \cite{Joshi1}. Here, we point out 
a few complementary issues that might be of observational interest. 

\begin{figure}[h!]
 \centering
 \subfigure[]{
 \includegraphics[width=2.5in,height=2.3in]{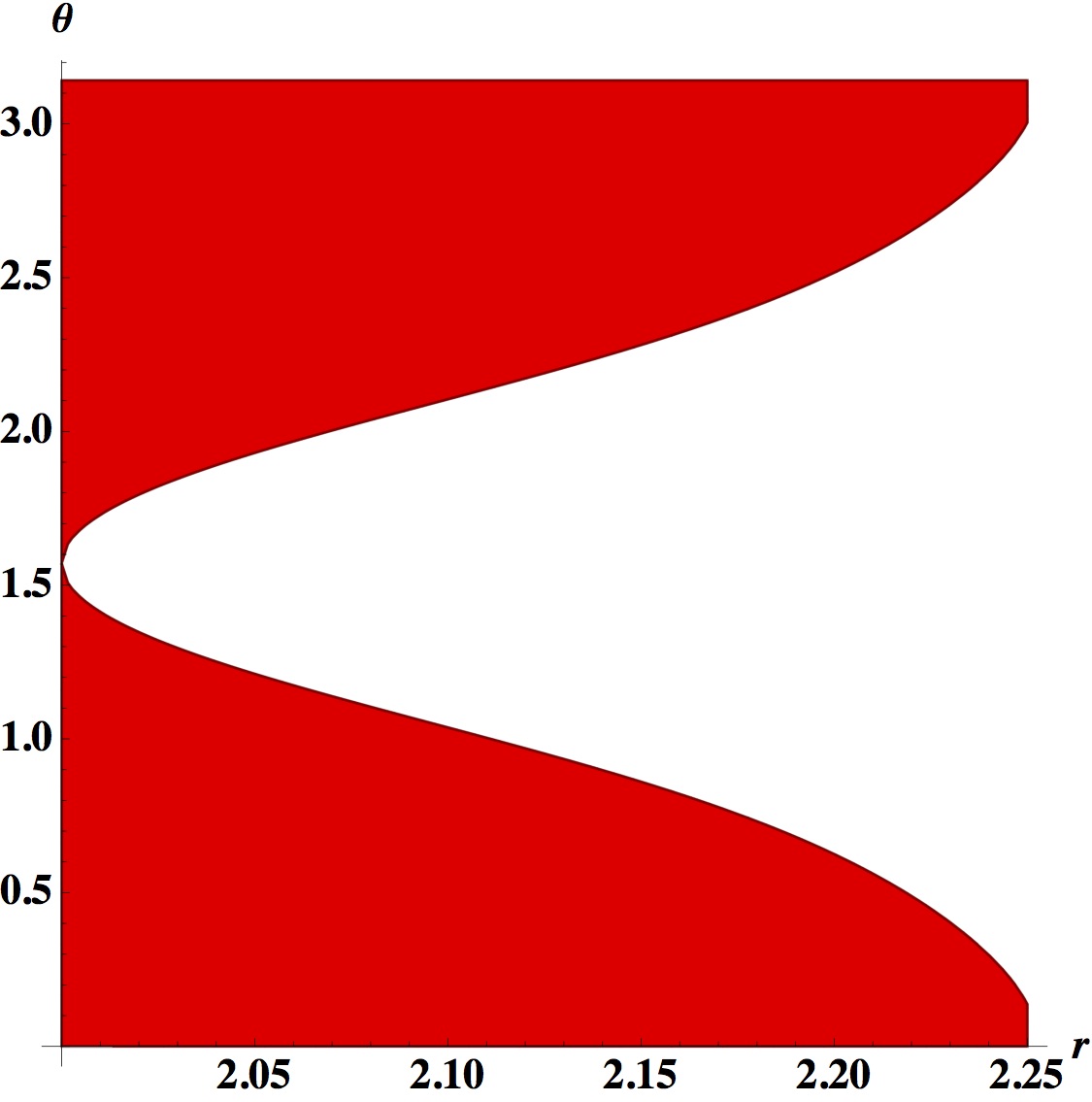}
 \label{LTnega} } 
 \subfigure[]{
 \includegraphics[width=2.5in,height=2.3in]{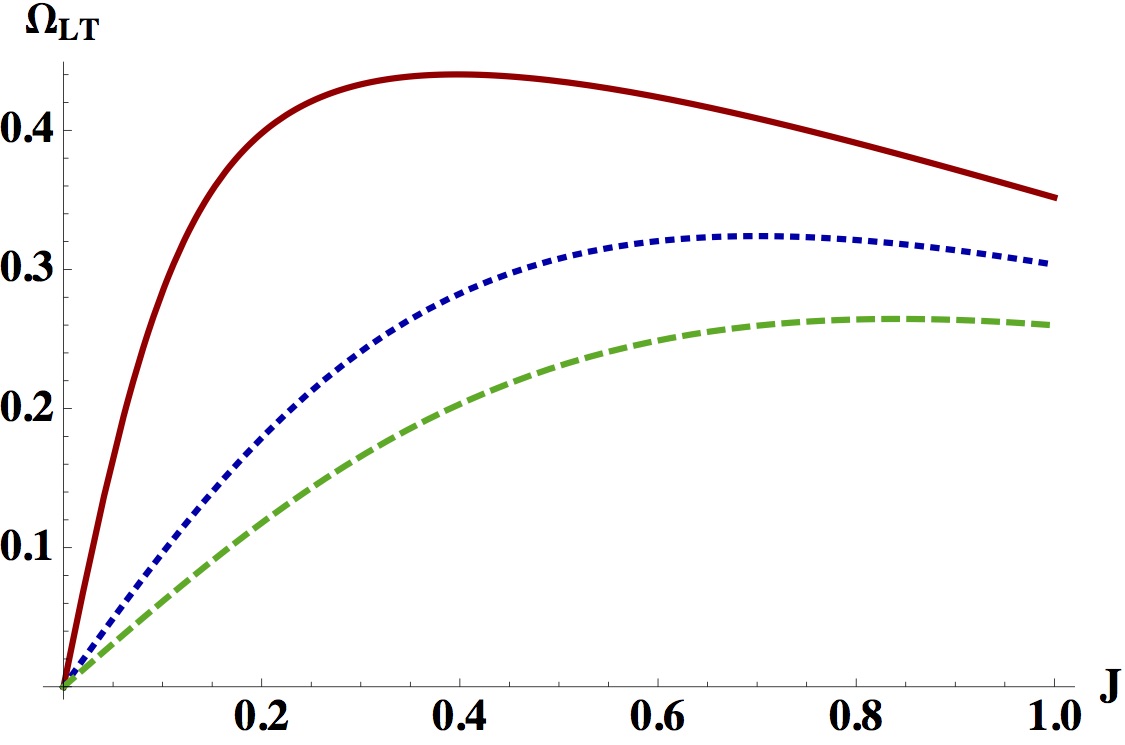}
 \label{LTnegb} }
  \caption{\small{(a)\, The red shaded region is the allowed region for $r$ and $\theta$ for which the right hand side of 
Eq.(\ref{LTder}) is negative, with $M=1$, $a=0.9$. (b)\,$\Omega^{\theta=0}_{LT}$ as a function of $J$ with $M=1$, for 
various values of $r=2.01$ (solid red), $r=2.1$ (dotted blue) and $r=2.2$ (dashed green)}}
 \label{LTneg}
 \end{figure}
First of all, we present the LT freq for $\theta = 0$ and $\theta = \pi/2$, for which Eq.(\ref{LTKerr}) is greatly simplified :
\begin{eqnarray}
\Omega_{LT}^{\theta = 0} = \frac{2 a M r}{\left(a^2+r^2\right)^{3/2} \sqrt{a^2+r (r-2 M)}}~,~~
\Omega_{LT}^{\theta = \pi/2} = \frac{a M}{r^2 (r-2 M)}
\label{LTspl}
\end{eqnarray}
From Eq.(\ref{LTspl}), we note that for $\theta = \pi/2$, i.e on the equitorial plane, the L-T precession frequency is a monotonically
increasing function of the angular momentum parameter, i.e for a given mass $M$, the frame dragging frequency becomes higher with
increasing angular momentum, which is physically expected. However, for $\theta = 0$ (and for all intermediate values of $\theta$ between
$0$ and $|\pi/2|$), 
the situation is qualitatively different. In particular, we have, 
\begin{equation}
\left(\frac{d\Omega_{LT}}{da}\right)_{\theta = 0} = \frac{r^3 (r-2M) -2 M r \left(3 a^4+2 a^2 r (r-2 M)\right)}
{\left(a^2+r^2\right)^{5/2} \left(a^2+r (r-2 M)\right)^{3/2}}~.
\label{LTder}
\end{equation}
For $r > 2M$, this quantity can be negative for some values of $r$. This is somewhat counter intuitive, as it indicates that (at least for
a fixed mass black hole), an increase in the rotation parameter might lead to a slow down in the L-T precession frequency. 

Our focus will be on regions outside the outer ergosurface. As we see below, this is enough for getting evidence on interesting physics. 
For specificness, we choose $M=1,~a=0.9$. Then, we will look for regions outside $r=2$ (the value of $r$ that will always be outside
the ergoregion for $M=1$ for any value of the rotation parameter) for which the right hand side of 
Eq.(\ref{LTder}) becomes negative. We show this as a region plot in Fig.(\ref{LTnega}), where the red shaded region satisfies this condition.
In Fig.(\ref{LTnegb}), we have depicted the result for $\theta=0$ for some specific choice of $r=2.01$ 
(solid red), $r=2.1$ (dotted blue) and $r=2.2$ (dashed
green) to show explicitly that this is the case. 
\begin{figure}[h!]
 \centering
 \subfigure[]{
 \includegraphics[width=2.5in,height=2.3in]{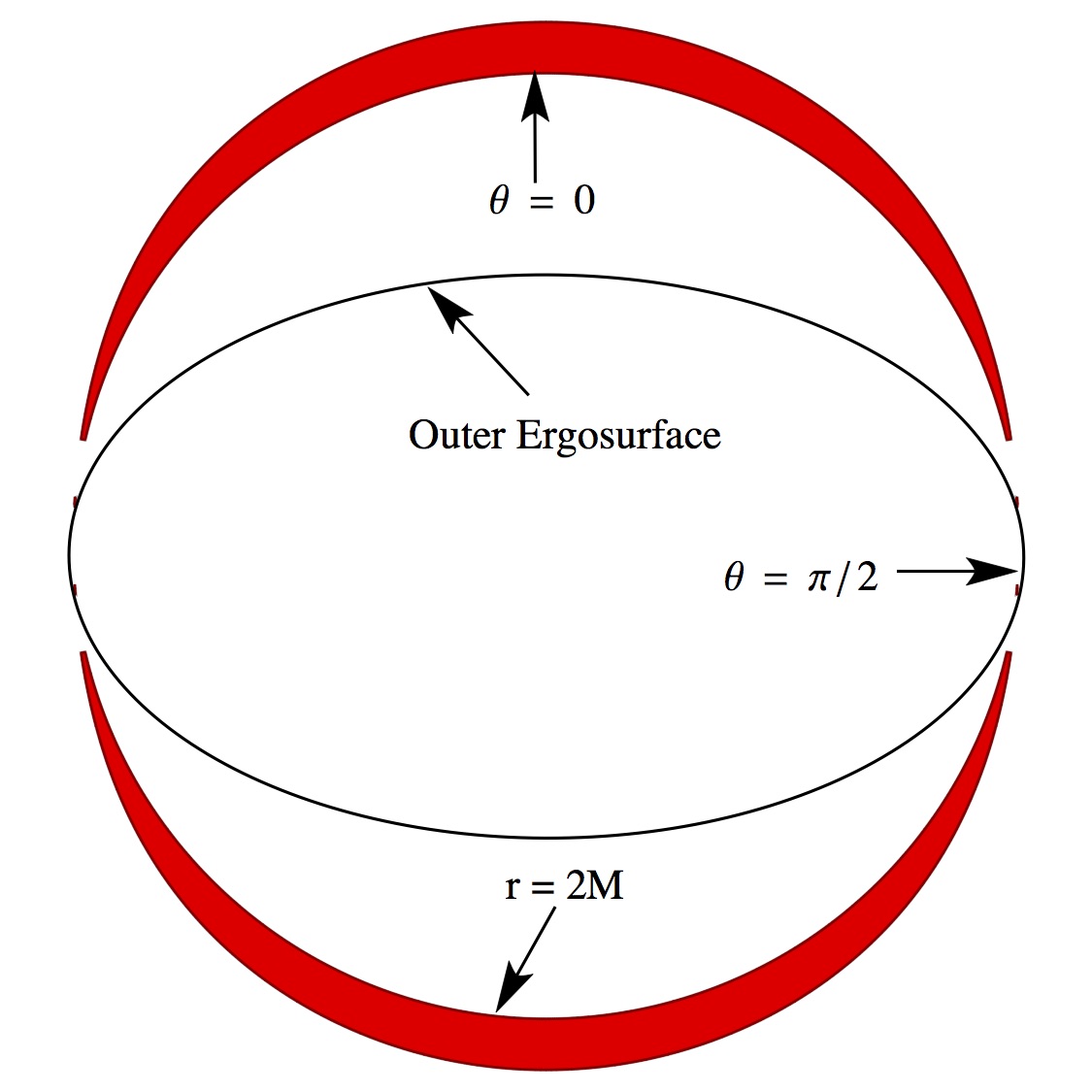}
 \label{Polara} } 
 \subfigure[]{
 \includegraphics[width=2.5in,height=2.3in]{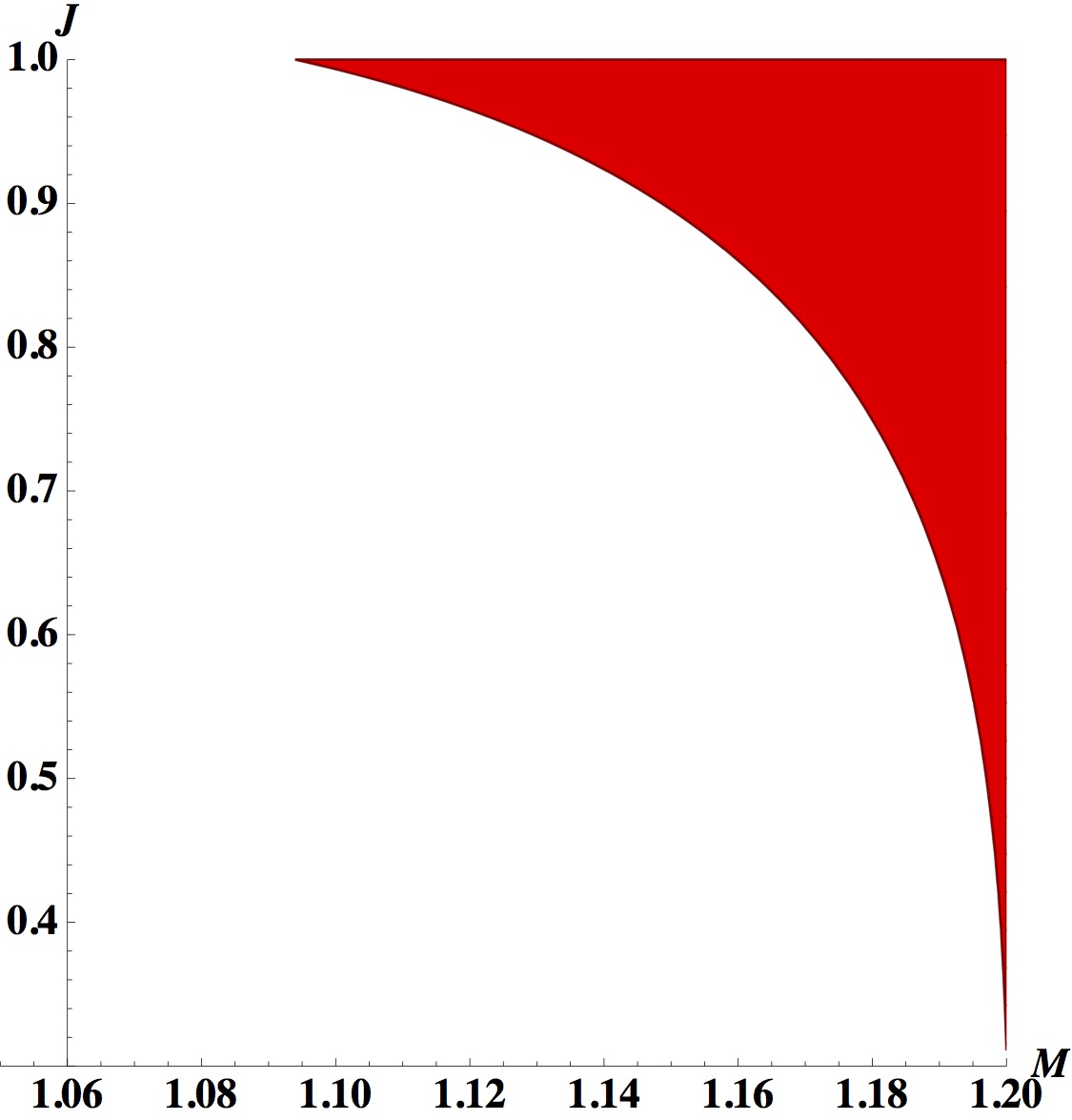}
 \label{Penrosea} }
  \caption{\small{(a)\, The red shaded region is the allowed region for $r$ and $\theta$ for which the right hand side of 
Eq.(\ref{LTder}) is negative, with $M=1.5$, $a=0.99$. (b)\, The red shaded region is the allowed region for $M$ and $J$ for which the right hand side of Eq.(\ref{LTder}) is negative, with $r=2.4$, $\theta=0$.}}
 \end{figure}

To understand the situation better, we have plotted in Fig.(\ref{Polara}), with $M = 1.5$ and $a=0.99$, the region in the $(r,\theta)$ plane where
the L-T frequency has a negative derivative. In this figure, we have searched for $d\Omega_{LT}/da <0$ starting from 
$r=2.3$, i.e outside the ergoregion. This plot is similar in spirit to the one in Fig.(\ref{LTnega}) -- one can see that starting from $\theta =0$, 
there is a region where this inequality is satisfied, all the way up to $|\theta| < \pi/2$. If one chooses smaller values of $a$, there are some 
unimportant variations in this, but the essential result is the same -- some part of the region outside the outer ergosurface shows a slowing
down of the L-T frequency when the rotation parameter is increased. 

Of course the above is for a fixed value of the black hole mass, and although it is reasonable to theoretically tune the angular momentum parameter
while keeping the mass fixed, one might wonder about the physicality of the result. To see this, in Fig.(\ref{Penrosea}), we have made a
region plot of the condition $d\Omega_{LT}/da <0$ with a fixed
$r=2.4$ and $\theta = 0$. We have varied the black hole mass from $M=1$ to $M=1.2$, so that our chosen value of $r$ always lies outside the
ergoregion, and we have taken $0<J<1$ so that a naked singularity is never formed. We see that there is indeed a region of parameters in 
the $(a,M)$ plane where the L-T frequency slows down. This is important for us, as we now explain. 

Slowing down of the L-T frequency with increasing $a$ is also equivalent to its speeding up with decreasing $a$. The latter phenomenon can
be achieved by the Penrose process, wherein one can extract energy out of a Kerr black hole that leads to a decrease of its angular momentum
as well as mass. This process specifically requires \cite{Townsend}
\begin{equation}
\frac{\delta J}{\delta M} \leq \frac{2M}{J}\left(M^2 + \sqrt{M^4 - J^2}\right)
\label{Pencon}
\end{equation}
From Fig.(\ref{Penrosea}), we can see that this can always be arranged deep inside the shaded red part. Thus, in principle, we conclude
that the slowing down of the L-T frequency can occur during a Penrose process. 

It is important to compare this with the accreting mass spinning black hole, which was considered several decades back by Bardeen \cite{BardeenKerr}.
There, it was shown under certain assumptions that due to the accretion process, the black hole angular momentum increases, and starting from
a Schwarzschild mass $M_1$, the value of the angular momentum per unit mass is \cite{BardeenKerr}
\begin{equation}
a =  \sqrt{\frac{2}{3}}M_1\left(4 - \sqrt{\frac{18M_1^2}{M^2} - 2}\right)
\label{Bard}
\end{equation}
with $0<a/M<1$ and $M_1\leq M<\sqrt{6}M_1$. Without loss of generality, we will set $M_1=1$ and we input this value of $a$ in
Eq.(\ref{LTKerr}). We choose $\theta = 0$ and $r \geq 4.9$ so that we are again outside the ergoregion. 
\begin{figure}[h!]
 \centering
 \subfigure[]{
 \includegraphics[width=2.5in,height=2.3in]{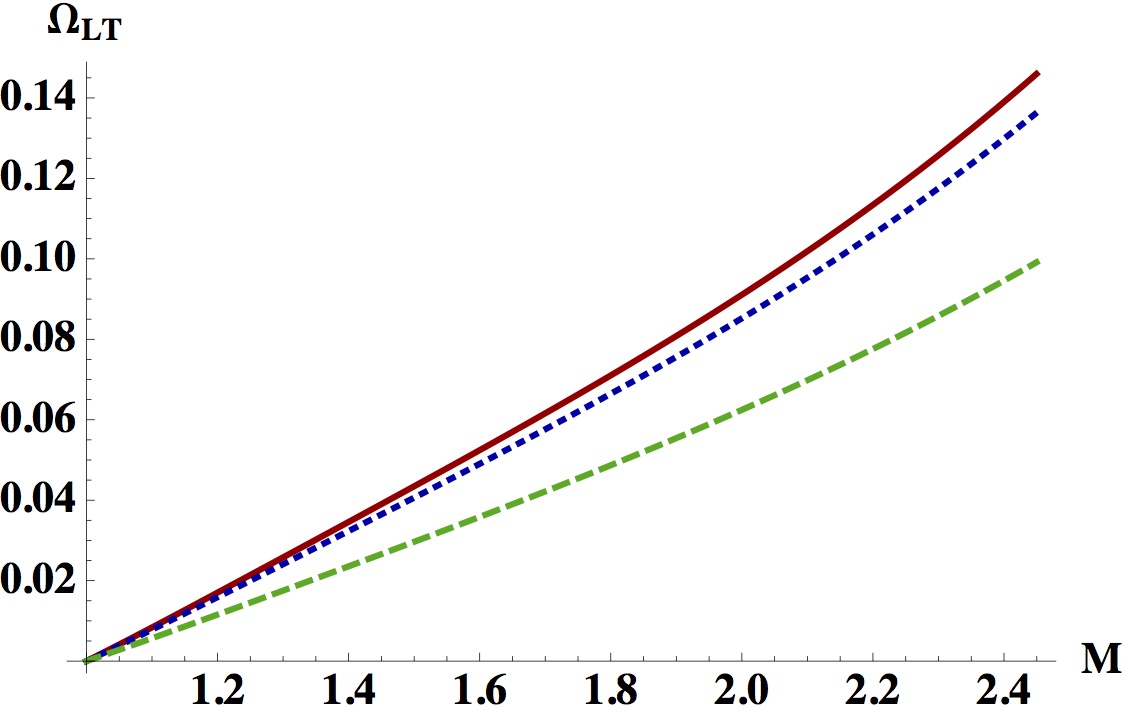}
 \label{acc1} } 
 \subfigure[]{
 \includegraphics[width=2.5in,height=2.3in]{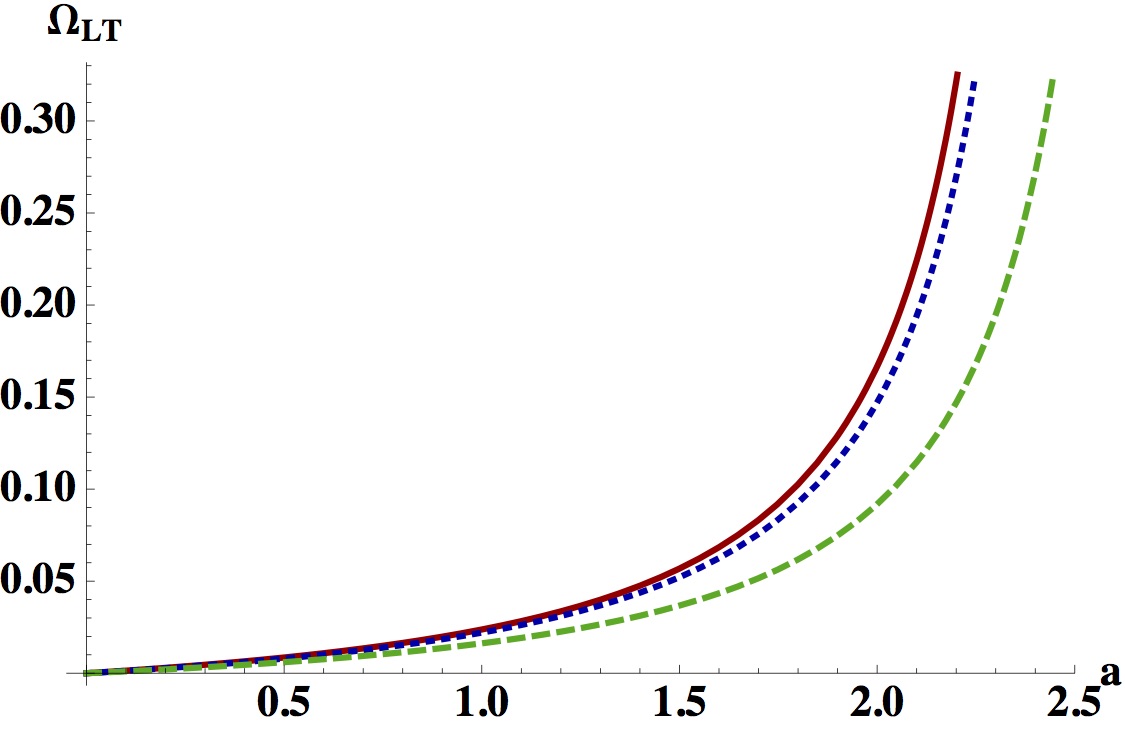}
 \label{acc2} }
  \caption{\small{(a)\, $\Omega_{LT}$ vs $M$ for $r=4.9$ (solid red), $r=5.0$ (dotted blue) and $r=5.5$ (dashed green) with $\theta = 0$. 
  (b)\, $\Omega_{LT}$ vs $M$ for $r=5.02$ (solid red), $r=5.1$ (dotted blue) and $r=5.5$ (dashed green) with $\theta = \pi/2$. Both these
  plots are for accreting Kerr black holes.}}
 \end{figure}
Fig.(\ref{acc1}) shows the result in this case. $\Omega_{LT}$ shows expected behaviour, it is linear in the weak field (small $M$) limit,
and non-linearity develops as $M$ becomes large. 
Here we show $\Omega_{LT}$ vs $M$ for $r=4.9$ (solid red), $r=5.0$ (dotted blue) and $r=5.5$ (dashed green) (with $\theta = 0$).
In Fig.(\ref{acc2}), we have shown the similar situation for $\theta = \pi/2$,
for completeness. Here, we have plotted $\Omega_{LT}$ as a function of $a$ after inverting Eq.(\ref{Bard}) and substituting
in Eq.(\ref{LTKerr}). In this plot, we have shown 
$\Omega_{LT}$ vs $M$ for $r=5.02$ (solid red), $r=5.1$ (dotted blue) and $r=5.5$ (dashed green) (with $\theta = \pi/2$) with the
value of $r$ chosen such that we are always outside the ergoregion.

We therefore see that in principle, the precession of a test gyro might speed up (slow down) in Kerr backgrounds, as the  
black hole spin parameter decreases (increases). The slowing down of the black hole might be ensured through a Penrose
process. However, for accreting Kerr black holes, this phenomenon is absent. 

\subsection{Strong field precession for generic stationary Kerr observer}

It is well known that inside the ergoregion of a Kerr black hole, it is not possible to define static observers. Stationary observers
can be defined both inside and outside the ergoregion of such black holes, and we will now concentrate on a class of stationary
observers outside the ergoregion, with their four velocities defined from the second of eq.(\ref{fourvels}). 
The precession frequency of test gyroscopes for such observers was calculated in 
\cite{JoshiLT2} and is given by a lengthy expression (eq.(14) of that paper) which follows readily from eq.(\ref{LTMain}) 
that we will not reproduce here. We will rather work in some limits. It turns out that the simplest
algebraic expressions are obtained in the equatorial plane $\theta = \pi/2$, which will be useful to decipher interesting physics in
an analytic manner. For other values of $\theta$, including $\theta = 0$, 
the expressions become cumbersome, and we will mostly deal with the equatorial plane, although we will make some comments
on other possible values of $\theta$ towards the end of this subsection. 

In the equatorial plane, the magnitude of the precession frequency (recall that this is the total precession frequency due to the
L-T precession and the de-Sitter effect, etc.) equals
\begin{equation}
\Omega_{F}^{\theta = \pi/2} = 
\frac{a^3 M \Omega ^2-2 a^2 M \Omega +a \left(3 M r^2 \Omega ^2+M\right)+r^2 \Omega
    (r-3 M)}{r^2 \left(r \left(a^2 \Omega ^2+r^2 \Omega ^2-1\right)+2 M (a \Omega
   -1)^2\right)}
\label{LTFthetapiby2}
\end{equation}
with the subscript ($F$) denoting that this is the total precession frequency. 
Let us first consider relatively large values of the radial coordinate, $r \gg 2M$. In this case, a series expansion yields,
\begin{equation}
\Omega_{F}^{\theta = \pi/2} \sim \frac{1}{\Omega r^2}  + \frac{1}{r^3}\left(3aM - \frac{3M}{\Omega}\right) 
+ {\mathcal O}\left(\frac{1}{r^4}\right)
\end{equation}
For this to be exactly zero to this order of approximation, 
we require that $\Omega = (3M-r)/(3aM)$. For large $r \gg 2M$, this will imply a large value of $\Omega$, since
$a$ lies between $0$ and $M$. However, $\Omega$ is restricted. It is well known (see, e.g \cite{Sorge},\cite{JoshiLT2}) that the 
allowed values of $\Omega$ for a time-like observer are restricted to lie between $\Omega^{+}$ and $\Omega^-$, where
\begin{equation}
\Omega^{\pm} = -\frac{g_{0\phi}}{g_{\phi\phi}} \pm \sqrt{\left(\frac{g_{0\phi}}{g_{\phi\phi}}\right)^2 - \frac{g_{0\phi}}{g_{\phi\phi}}}
\end{equation}
Using the metric components from eq.(\ref{Kerrmetric}), we find that in the limit of large $r$, on the equatorial plane, 
\begin{equation}
\Omega^{\pm} \sim \pm\frac{1}{r} \mp \frac{M}{r^2} \mp \frac{a^2 \mp 4aM + M^2}{2r^3}+ {\mathcal O}\left(\frac{1}{r^4}\right)
\end{equation}
Hence, we conclude that the precession is non-zero for relatively large values $r$, with its sign being determined by that of $\Omega$.
Of course strictly in the limit $r\to \infty$, $\Omega_{F}^{\theta = \pi/2}$ vanishes, as it should. 

On the other hand, we note that in general, setting $\Omega_{F}^{\theta = \pi/2} = 0$\footnote{This case is physically relevant as it
yields a precise vanishing of the gyro precession frequency, without any approximation.} yields two special values $\Omega^{\theta=\pi/2}_{0\pm}$, i.e
\begin{equation}
\Omega^{\theta=\pi/2}_{0\pm} = 
\frac{2 a^2 M+3 M r^2-r^3 \pm r^{3/2}\sqrt{\left(r (3 M-r)^2-4 a^2 M\right)}}{2 \left(a^3
   M+3 a M r^2\right)}
\label{impzero}
\end{equation}
with the subscript $0$ denoting that this value of the angular velocity of the stationary observer makes the precession frequency
$\Omega_F$ vanish exactly. Apart from $r=3M$, the quantity in the square root can always be made positive so that 
$\Omega^{\theta=\pi/2}_{0\pm}$ is real. 
Near $r \to 2M$, we have
\begin{equation}
\Omega^{\theta=\pi/2}_{0\pm} \to \frac{a^2 M+2 M^3 \pm 2 M^2\sqrt{M^2-2 a^2}}{a^3 M+12 a M^3},~~{\rm as}~~
r \to 2M
\end{equation}
This shows that in units of $M$, close to the outer ergosphere $r=2M$, which is of our interest here,
$\Omega^{\theta=\pi/2}_{0\pm}$ is real for $a < M/\sqrt{2}$. 
It can be checked that in this region, $\Omega^{\theta=\pi/2}_{0-}$ can 
certainly lie in the range $\Omega^+ > \Omega^{\theta=\pi/2}_{0-}>\Omega^-$, 
with $\theta = \pi/2$, and that $\Omega^{\theta=\pi/2}_{0+}$ will in general not satisfy this contraint. 
Henceforth, to avoid cluttering of notation, we call $\Omega^{\theta=\pi/2}_{0-}=\Omega^{\theta=\pi/2}_{0}$.
\begin{figure}[h!]
 \centering
 \subfigure[]{
 \includegraphics[width=2.5in,height=2.3in]{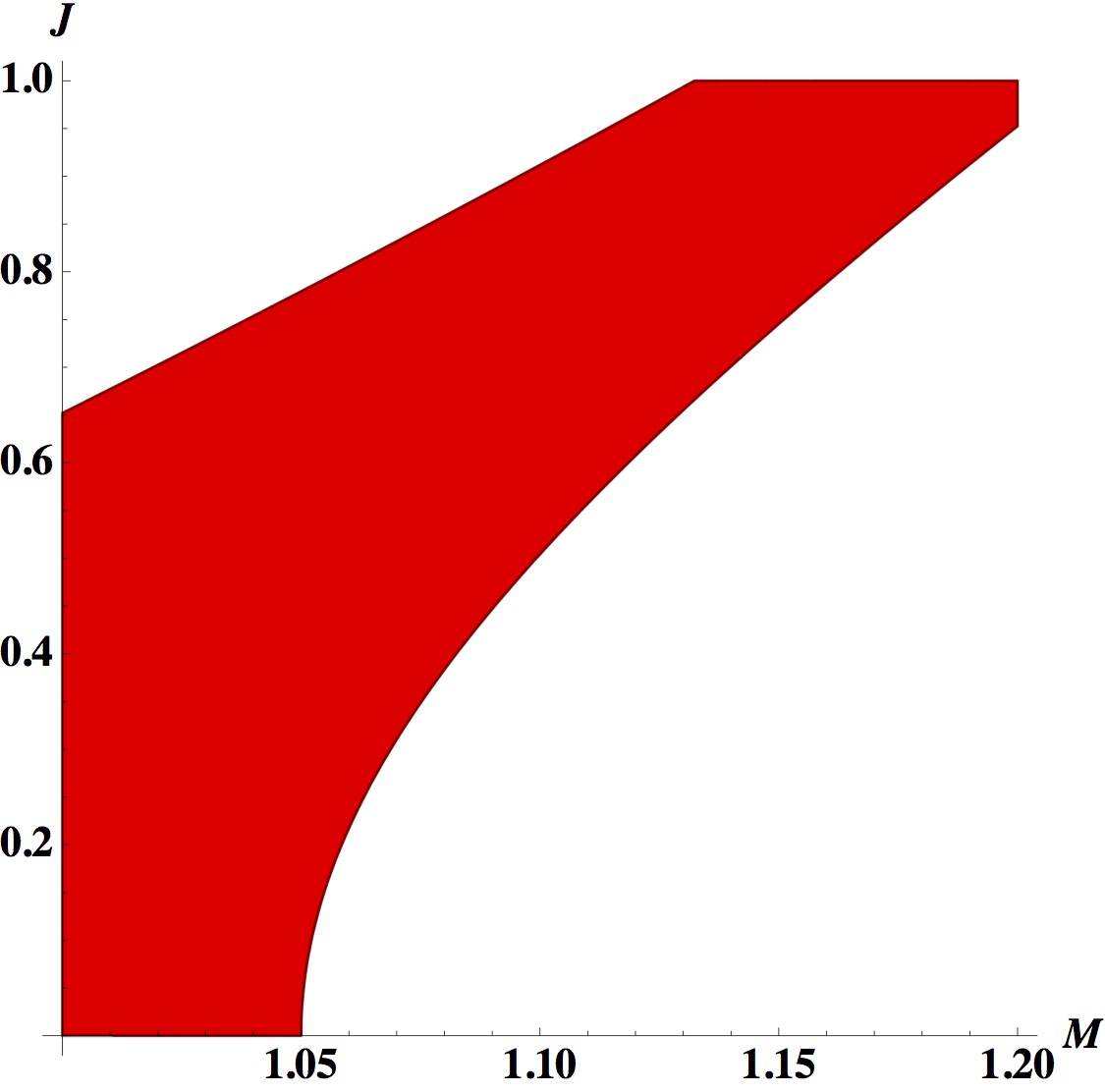}
 \label{zerofreq1} } 
 \subfigure[]{
 \includegraphics[width=2.5in,height=2.3in]{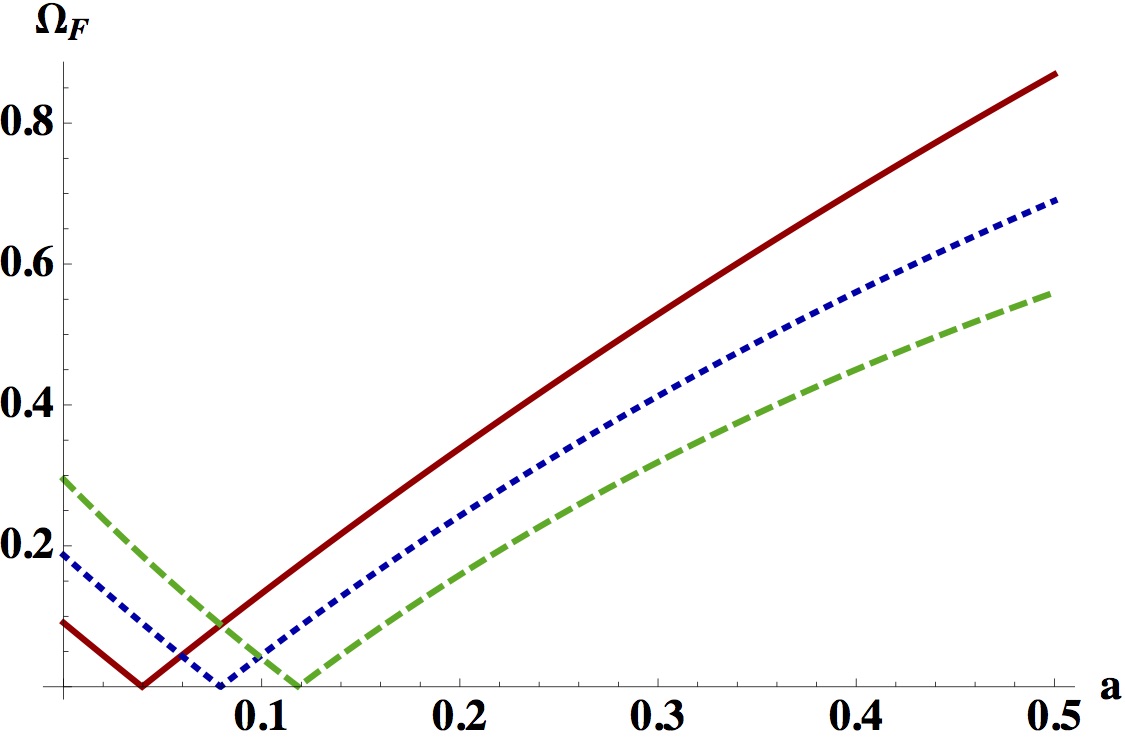}
 \label{zerofreq1a} }
  \caption{\small{(a)\, Region plot (red shaded region) of the condition $\Omega^+>\Omega^{\theta=\pi/2}_{0}>\Omega^-$ for $r=2.1$. 
  (b)\,  Precession frequency vs the rotation parameter for $M=1$, $r=2.1$. The solid red, dotted blue and dashed green curves correspond to
$\Omega = 0.01$, $0.02$ and $0.03$, respectively.}}
 \end{figure}

As a concrete illustration, we choose $M=1$. Then, in units of $M$, we take $r>2$ so
that we are outside the ergoregion. A region plot (red shaded region) of the condition $\Omega^+>\Omega^{\theta=\pi/2}_{0}>\Omega^-$ is 
shown in fig.(\ref{zerofreq1}) in terms of $M$ and $J$ with the radial coordinate $r=2.1$. 
In fig.(\ref{zerofreq1a}), we show some explicit plots of the gyro precession frequency $\Omega_F$ vs the rotation parameter $a$. Here
again we have taken $M=1$, and have again chosen $r=2.1$. The solid red, dotted blue and dashed green curves 
correspond to $\Omega = 0.01$, $0.02$ and $0.03$, respectively (For this choice of $M$ and $r$, $\Omega^+$ varies between $0.1$ 
($J=0$) to $0.32$ ($J=1$), and our choices of $\Omega$ are below this. The values of $a \equiv J$ that these correspond to, at $\Omega_F=0$
are seen to be in the region depicted in fig.(\ref{zerofreq1}). 

\begin{figure}[h!]
 \centering
 \subfigure[]{
 \includegraphics[width=2.5in,height=2.3in]{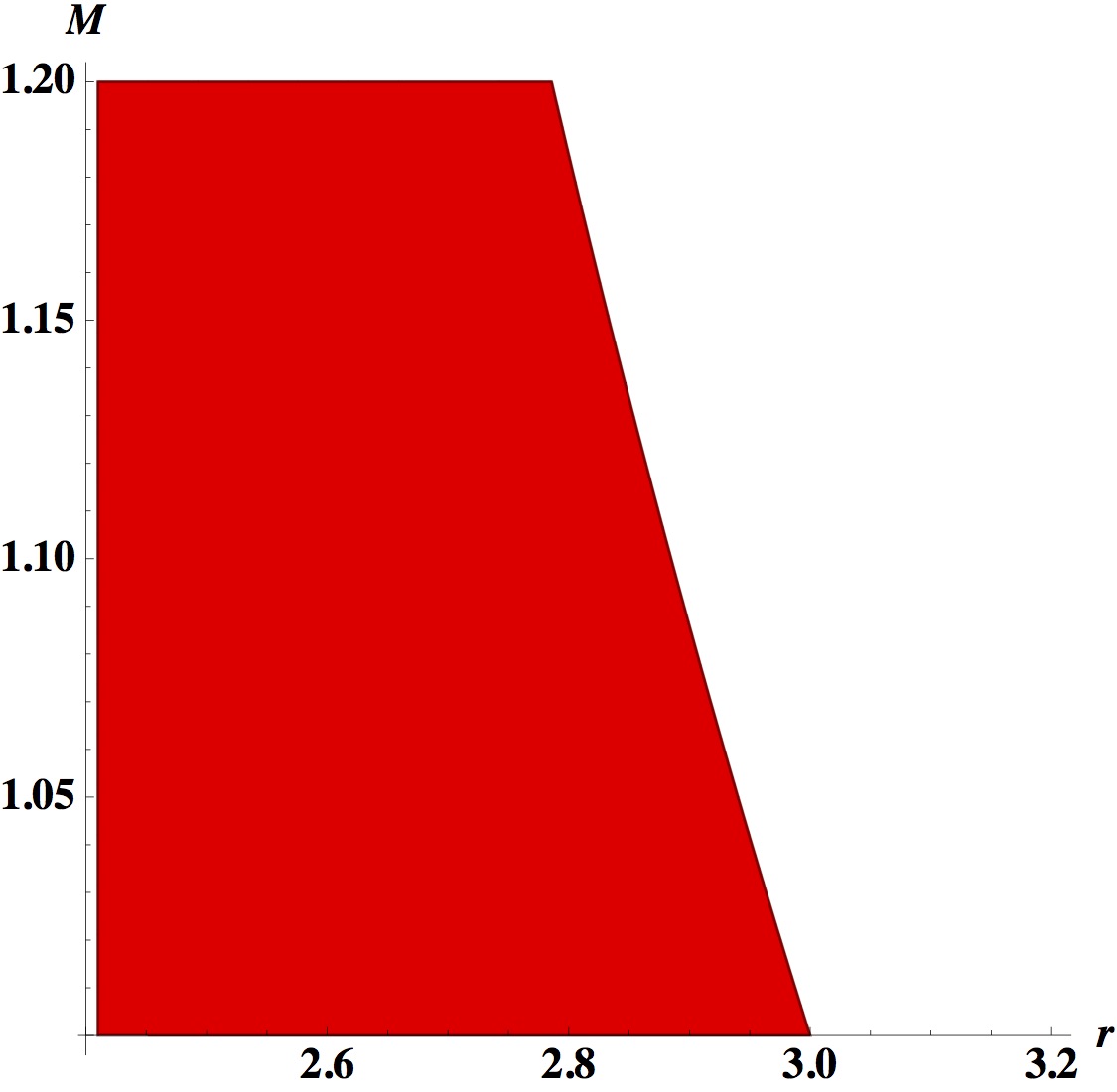}
 \label{zerofreq2} } 
 \subfigure[]{
 \includegraphics[width=2.5in,height=2.3in]{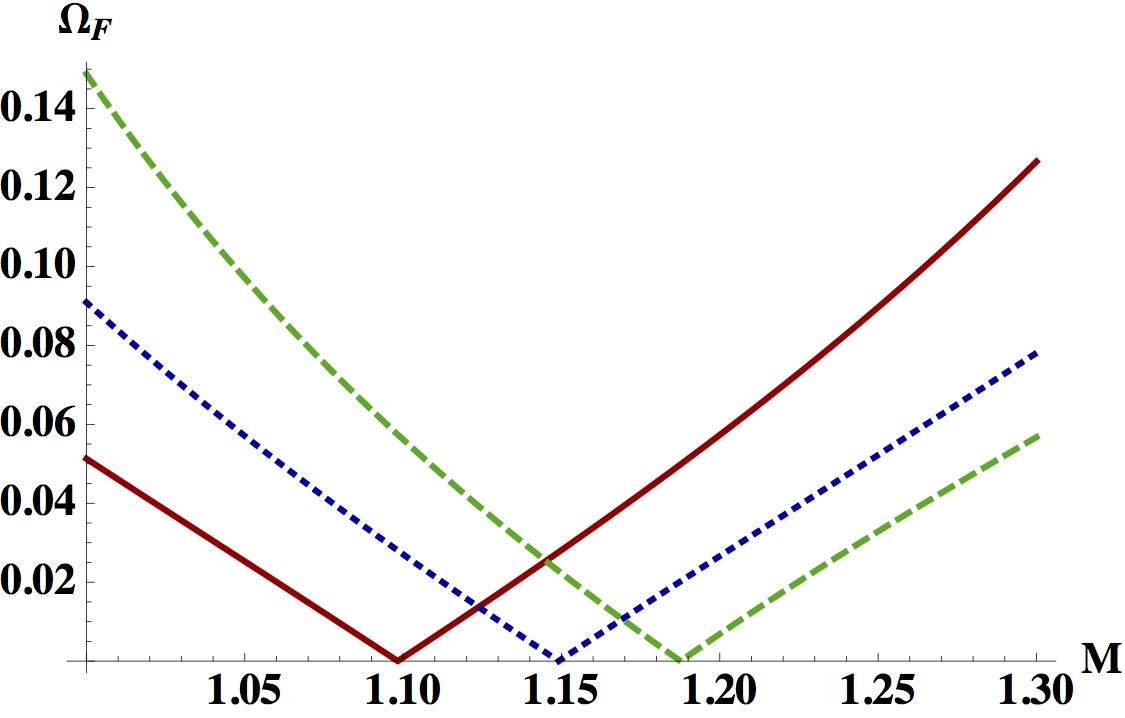}
 \label{zerofreq2a} }
  \caption{\small{(a)\, Region plot (red shaded region) of the condition $\Omega^+>\Omega^{\theta=\pi/2}_{0}>\Omega^-$ with $a$
 related to $M$ by eq.(\ref{impzero}). 
  (b)\,  Precession frequency vs $M$ for the accreting mass Kerr black hole, for $r=2.73$. The solid red, dotted blue and dashed 
  green curves correspond to
$\Omega = 0.1$, $0.13$ and $0.15$, respectively.}}
 \end{figure}

We now turn to a similar analysis for accreting Kerr black holes, and substitute the value of $a$ from eq.(\ref{Bard}) in 
eq.(\ref{impzero}). The condition that  $\Omega^+>\Omega^{\theta=\pi/2}_{0}>\Omega^-$ is plotted as the red shaded region
in the $(r,M)$ plane in fig.(\ref{zerofreq2}) (where we start from $M=1,~r=2$ to avoid the ergoregion) and explicit plots of
$\Omega_F$ as a function of $M$ for $r=2.73$ is shown in fig.(\ref{zerofreq2a}), where the solid red, dotted blue and the
dashed green curves correspond to $\Omega = 0.1$, $0.13$ and $0.15$, respectively. For this case, we see that the precession
frequency of a test gyroscope can indeed vanish for a Copernican stationary observer for the accreting black hole. Although analytical
results are cumbersome to produce here, we note that one can numerically verify the vanishing of the precession frequency, and
confirm this with the predictions of fig.(\ref{zerofreq2a}). 

The above considerations point to the fact that for a Kerr black hole, with change of the angular momentum parameter, Copernican
observers at a certain radius outside the black hole will see that the L-T precession frequency in their frame will change sign. The
same is true for accreting Kerr black holes as well. Whether this has any experimental relevance is not immediately clear, but should
be an interesting topic to probe further. 

Finally, a few words about values of $\theta \neq \pi/2$. Let us consider the case $\theta = 0$. In this case, we find the magnitude
of the L-T precession frequency of the Kerr black hole as
\begin{eqnarray}
&&\Omega_{F}^{\theta = 0} =\frac{\sqrt{{\mathcal A^2} + {\mathcal B^2}}}{\left(a^2+r^2\right)^{3/2}\left(a^2 + r(r-2M)\right)},\nonumber\\
&&{\mathcal A} = \Omega  \left(a^2+r^2\right) \left(a^2 (M+r)+r^2 (r-3 M)\right),\nonumber\\
&&{\mathcal B}= \sqrt{a^2+r (r-2 M)} \left(\Omega  \left(a^2+r^2\right)^2-2 a M r\right)
\end{eqnarray}
This expression is manifestly non-zero for all values of $\Omega$. It therefore seems that the exact vanishing of the gyroscope 
precession frequency for the Kerr black hole is possibly manifest only on the equatorial plane. We have numerically 
checked for some other values of $\theta$ that this last statement is true. 

\subsection{Strong field Lense-Thirring precession for static observers : rotating JNW solution}

We will now discuss the strong field L-T precession for static observers in rotating JNW backgrounds. 
We start with the metric derived in \cite{KroriBhattacharjee} (see \cite{Gyulchev}, \cite{Harko} for recent works on this).
The metric  of the rotating Janis-Newman-Winicour naked singularity in Boyer-Lindquist coordinates is given as :
\begin{eqnarray}
ds^2 &=& - \left(1 - \frac{2Mr}{\gamma \rho^2}\right)^\gamma \left(dt-\omega d\phi \right)^2 +
\left(1 - \frac{2Mr}{\gamma \rho^2}\right)^{1-\gamma} \rho^2\left(\frac{dr^2}{\Delta_1}+d\theta^2+\sin^2\theta d\phi^2\right)\nonumber\\
&-& 2w(dt-\omega d\phi)d\phi
\label{eq1}
\end{eqnarray}
where we have defined
\begin{equation}
\omega=a\sin^2\theta~,~\rho^2=r^2+a^2\cos^2\theta~,~\Delta_1=r^2+a^2-\frac{2Mr}{\gamma}
\label{eq2}
\end{equation}
This space-time is sourced by the scalar field $\psi$ satisfying $\Box \psi = 0$ in this background, and is given by
\begin{equation}
\psi=\frac{1}{4}\sqrt{1-\gamma^2}\log\left(1-\frac{2Mr}{\gamma\rho^2}\right)
\end{equation}
The deformation parameter $\gamma$ is always positive and lies between $0$ and $1$. 

Here, $M$ is the ADM mass, and $a$ is as before, the 
angular momentum per unit mass of the source. We remind the reader that as before, we have assumed units with $G=c=1$.
The metric of eq.(\ref{eq1}) reduces to the Kerr solution with $\gamma = 1$, to the non-rotating JNW solution with $a=0$
and to the Schwarzschild solution with both $\gamma=1$ and $a=0$. Calculating the Ricci scalar, one finds that there is
a globally naked singularity at $g_{tt}=0$, which translates into 
\begin{equation}
r_{\pm} = \frac{M}{\gamma} \pm\sqrt{\left(\frac{M}{\gamma}\right)^2 -a^2\cos^2\theta}
\label{NSlocation}
\end{equation}
The surface $r=r_+$ is the naked singularity, which occurs at $r_+ = 2M/\gamma$ for $\theta=\pi/2$.
Also, from the above equation it is seen that here the angular momentum parameter
$a$ can range from $0$ to $M/\gamma$, for a given value of $\gamma$. Here, $\Delta_1=0$ yields 
\begin{equation}
r^{\Delta_1=0}_{\pm} = \frac{M}{\gamma} \pm\sqrt{\left(\frac{M}{\gamma}\right)^2 -a^2}
\end{equation}
so that the radial values $r^{\Delta_1=0}_{\pm}$ are always less than $r_+$ defined by eq.(\ref{NSlocation}), so that there
is no ergoregion. 

The magnitude of the strong field L-T precession can be readily calculated, and yields a complicated expression. We will first look at some limits.
In the limit $r \gg 2M$, we have from eq.(\ref{LTMain})\footnote{We will use the same symbol for the L-T precession as we did for Kerr 
backgrounds, in order not to clutter the notation. The meaning should be clear from the context.}
\begin{equation}
\Omega_{LT} = \frac{aM}{r^3}\sqrt{4\cos^2\theta + \sin^2\theta} + \frac{aM^2}{r^4\gamma}
\frac{4\cos^2\theta\left(3-2\gamma\right) + \sin^2\theta\left(3-\gamma\right)}{\sqrt{4\cos^2\theta + \sin^2\theta}} + {\mathcal O}
\left(\frac{1}{r^5}\right)
\label{weakJNW}
\end{equation}
We note that the first term is similar to the one in eq.(\ref{weak}) for the Kerr background, while the second term might
be important for large $r$ with small $\gamma$, where it might contribute at ${\mathcal O}(r^{-3})$.
Also, for $\theta = 0$ and $\theta = \pi/2$, we obtain
\begin{equation}
\Omega_{LT}^{\theta = 0} = \frac{a\gamma \left[\left(1-\frac{2 M r}{\gamma\left(a^2 + r^2\right)}\right)^{\gamma }-1\right]}
{\left[\left(a^2 + r^2\right)\gamma\right]^{\gamma/2}\left[\left(a^2+r^2\right)\gamma - 2Mr\right]^{1-\gamma/2}},~
\Omega_{LT}^{\theta = \pi/2} = \frac{a \gamma ^{3/2} M \left(1-\frac{2 M}{\gamma  r}\right)^{\gamma /2}}{r^{3/2}
(\gamma  r-2 M)^{3/2}}
\label{LTsplJNW}
\end{equation}
The expressions in eq.(\ref{LTsplJNW}) reduces to the corresponding ones for the Kerr metric in eq.(\ref{LTspl}) on 
setting $\gamma = 1$.
\begin{figure}[h!]
 \centering
 \subfigure[]{
 \includegraphics[width=2.5in,height=2.3in]{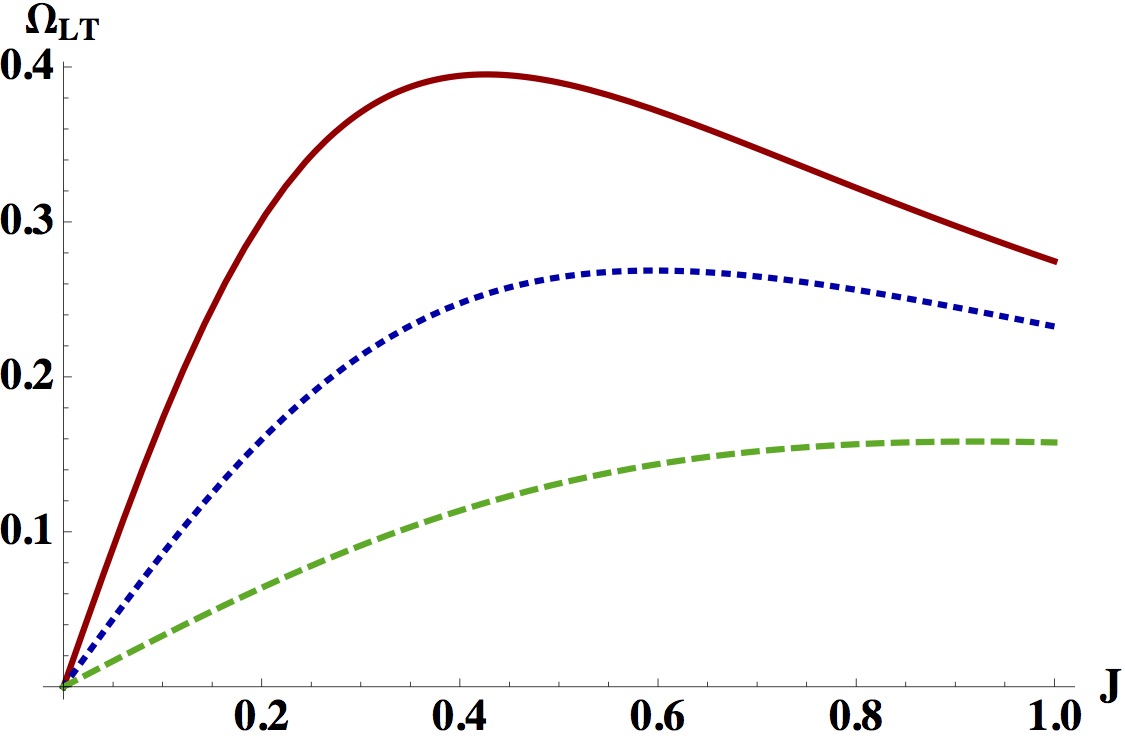}
 \label{JNW1} } 
 \subfigure[]{
 \includegraphics[width=2.5in,height=2.3in]{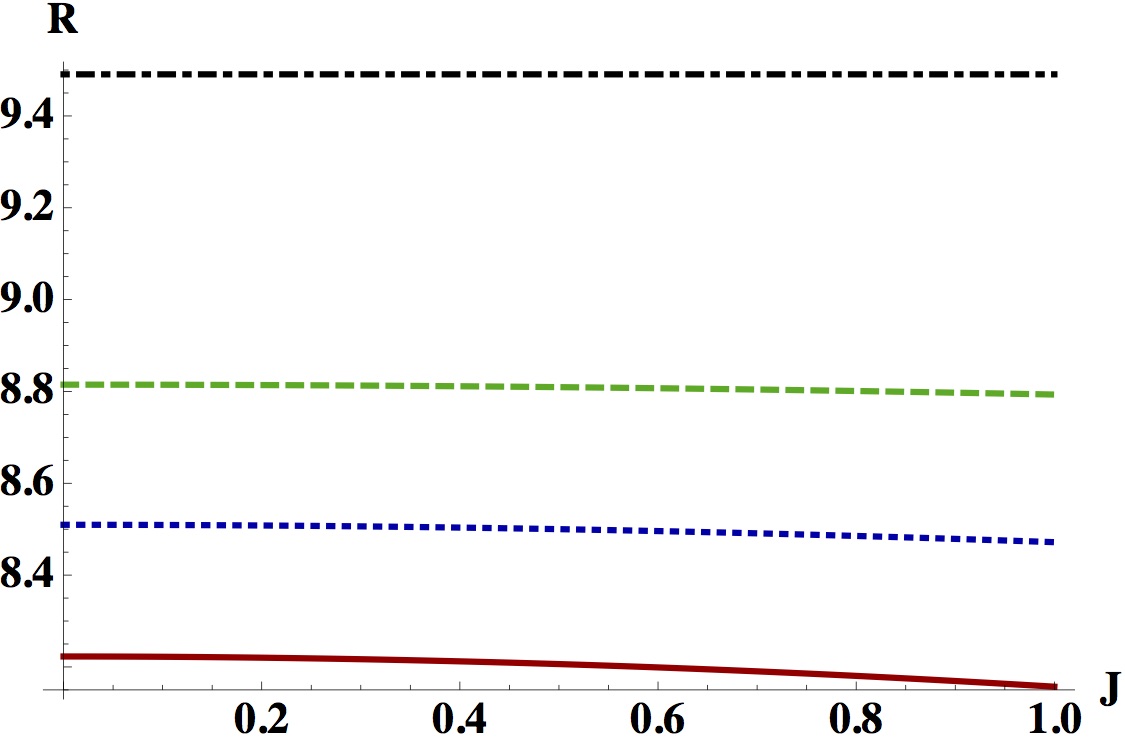}
 \label{JNW2} }
\caption{\small{(a)\, $\Omega^{\theta=0}_{LT}$ vs $J$ with $M=1$ and $\gamma=0.1$ in the JNW background. 
The solid red, dotted blue and dashed green 
curves correspond to $r=20.01$, $20.02$ and $20.05$, respectively.  
(b)\, The ratio $R$ of eq.(\ref{ratio}) plotted as a function of $J$ for $M=1$, $r=25$. The solid red, dotted blue, dashed green 
and dot-dashed black curves correspond to $\theta = 0$, $\pi/4$, $\pi/3$ and $\pi/2$, respectively.}}
 \end{figure}
For the rotating JNW singularity, analysis of the expressions for the L-T precession for generic values of $\theta$ become
difficult due to algebraic complications. However, we find that the qualitative features of the precession to 
remain the same as in the Kerr case, with the coordinates scaled. Remember that the singularity is located at 
\begin{equation}
r_{+} = \frac{M}{\gamma}+\sqrt{\left(\frac{M}{\gamma}\right)^2 - a^2\cos^2\theta}
\label{sing}
\end{equation}
Comparing with eq.(\ref{conds}), we see that smaller values of $\gamma$ imply that the singularity is located at 
relatively larger values of the radial coordinate as compared with the Kerr case. 

Our results are illustrated in fig.(\ref{JNW1}), where
we have shown the behaviour of  $\Omega^{\theta=0}_{LT}$ vs $J$ with $M=1$, and $\gamma=0.1$ in the JNW space-time. Here,
we have chosen the values of $r > 20$ to be outside the singular surface. 
Specifically, the solid red, dotted blue and dashed green curves correspond to $r=20.01$, $20.02$ and $20.05$, respectively. We note the
strong similarity with the corresponding Kerr case depicted in fig.(\ref{LTnegb}). Hence we conclude that the essential nature
of the the L-T precession frequency is the same just outside the naked singularity compared to Kerr black hole just outside the
ergoregion. 

More interesting physics can be gleaned by comparing the magnitudes of the L-T frequency for the JNW and Kerr space-times. 
We define a quantity
\begin{equation}
R = \frac{\Omega^{JNW}_{LT}}{\Omega^{Kerr}_{LT}}
\label{ratio}
\end{equation}
In fig.(\ref{JNW2}), we plot this quantity as a function of the angular momentum $J$, for $M=1$ and $r=25$, where for the
JNW space-time we have taken $\gamma = 0.1$. In this figure, the solid red, 
dotted blue, dashed green and dot-dashed black curves correspond to $\theta = 0$, $\pi/4$, $\pi/3$ and $\pi/2$, respectively.
We see that this ratio increases as one approaches the equatorial plane, where the L-T precession might be almost an order
of magnitude higher for the JNW naked singularity compared to the Kerr black hole. Strong enhancement of L-T precession
frequency might thus be a characteristic distinction between black holes and naked singularities. 

It is not difficult to undertake a similar analysis for JNW stationary observers, as we had done for the Kerr background, although 
the expressions become very complicated. However,
our discussion above gives indication that essentially similar features will persist in this case also, at different radial distances. 
This will possibly not lead to any significantly new physical insight, and we will not perform such an analysis here. 

We now move over to another aspect of importance in observational distinction between black holes and naked singularities, 
namely tidal effects. 

\section{Tidal forces in rotating JNW backgrounds}

In this section, we will compute tidal forces in rotating JNW backgrounds, using the metric of eq.(\ref{eq1}). The general 
analysis of arbitrary orbits is substantially complicated, and we will try to illustrate the essential physics by 
considering only equatorial circular orbits ($\theta=\frac{\pi}{2}$ and fixed $r$). 
By \cite{Ishii}, the tidal potential in Fermi normal coordinates is given by:
\begin{multline}
\phi_{tidal}=\frac{1}{2}C_{ij}x^ix^j+\frac{1}{6}C_{ijk}x^ix^jx^k+\frac{1}{24}[C_{ijkl}+4C_{(ij}C_{kl)}-4B_{(kl|n|}B_{ij)n}]x^ix^jx^kx^l
\label{eq3}
\end{multline}
Where we have defined
\begin{equation}
C_{ij} =R_{0i0j},~ C_{ijk}=R_{0(i|0|j;k)},~ C_{ijkl}=R_{0(i|0|j;kl)},~B_{ijk}=R_{k(ij)0},~A_{k}=\frac{2}{3}B_{ijk}x^ix^j
\label{eq4}
\end{equation}
Here $R_{0(i|m|j;kl)}$ means (in accordance with notation of \cite{Ishii}) sum over permutations of $i$, $j$, $k$ and $l$,
keeping $m$ at fixed position and then dividing by total number of permutations. All the components refer to Fermi normal coordinates. 
Our first task is therefore to compute the Fermi normal coordinates for equatorial circular orbits, from which we can explicitly
compute the tidal potential of eq.(\ref{eq3}) after plugging in the expressions from eq.(\ref{eq4}). The calculation can 
be tedious, and we will first propose a relatively simple method to compute Fermi normal coordinates (for equatorial circular orbits)
for a generic class of metrics written in Boyer-Lindquist coordinates.

\subsection{Equatorial circular orbits and Fermi normal coordinates}

Here we present a simple analysis to construct Fermi normal coordinates on equatorial circular geodesics for a certain class 
of metrics. Consider a general metric in Boyer-Lindquist coordinates:
\begin{equation}
ds^2 =  g_{tt}(r,\theta)dt^2+2g_{t\phi}(r,\theta)dtd\phi+g_{\phi\phi}(r,\theta)d\phi^2+g_{rr}(r,\theta)dr^2+g_{\theta\theta}(r,\theta)d\theta^2
\label{eq5}
\end{equation}
Here metric tensor components are dependent only on $\theta$ and $r$. Our convention is chosen such that at large $r$, 
the metric reduces to the Minkowski metric with $\eta_{\mu\nu}=diag\{-1,1,1,1\}$ (although this is not strictly necessary). 
The Lagrangian for a massive object on equatorial circular orbit is given by:
\begin{equation}
{\mathcal L}=\frac{1}{2}\left(g_{tt}\dot{t}^2+2g_{t\phi}\dot{t}\dot{\phi}+g_{\phi\phi}\dot{\phi}^2\right)
\label{eq6}
\end{equation}
Where dot denotes differentiation with respect to the proper time $\tau$. All metric components are evaluated for 
$\theta=\frac{\pi}{2}$. Now, $t$ and $\phi$ being cyclic coordinates, we get two constants of motion, namely the 
energy per unit mass $E$ and angular momentum per unit mass $L$. These are given by
\begin{equation}
E = -(g_{tt}\dot{t}+g_{t\phi}\dot{\phi}),~~
L = (g_{t\phi}\dot{t}+g_{\phi\phi}\dot{\phi})
\label{eq7}
\end{equation}
Solving for $\dot{t}$ and $\dot{\phi}$ gives 
\begin{equation}
\dot{t} = \frac{1}{g}(g_{t\phi}L+g_{\phi\phi}E),~~
\dot{\phi} = -\frac{1}{g}(g_{tt}L+g_{t\phi}E)
\label{eq8}
\end{equation}
Where we have denoted $g=g_{t\phi}^2-g_{tt}g_{\phi\phi}$. As the object is massive, we get a constraint on $E$ and $L$ 
from rest mass conservation,
\begin{equation}
g_{\phi\phi}E^2+2g_{t\phi}EL+g_{tt}L^2=g
\label{eq9}
\end{equation}
Moreover, for equatorial circular orbits, we have
\begin{equation}
\frac{\partial g_{\phi\phi}}{\partial r}E^2+2\frac{\partial g_{t\phi}}{\partial r}EL+\frac{\partial g_{tt}}{\partial r}L^2=\frac{\partial g}{\partial r}
\label{eq10}
\end{equation}
To construct Fermi normal coordinates we need one time-like and three space-like vectors which are parallely propagated along the
geodesic. As the tangent vector is always parallely propagated along a geodesic, we choose this as the time-like vector.
\begin{equation}
\lambda_{0}^{\mu}=\{\dot{t},0,0,\dot{\phi}\}
=\{\frac{1}{g}(g_{t\phi}L+g_{\phi\phi}E),0,0,-\frac{1}{g}(g_{tt}L+g_{t\phi}E)\}
\label{eq11}
\end{equation}
Also it is seen that for this kind of metric (being independent of $t$ and $\phi$), we get that
\begin{equation}
\lambda_{2}^{\mu}=\{0,0,\frac{1}{\sqrt{g_{\theta\theta}}},0\}\\
\label{eq12}
\end{equation}
is a space-like vector and is parallely propagated along a equatorial circular geodesic. Two obvious orthonormal space-like vectors are then
\begin{equation}
\widetilde{\lambda}_{1}^{\mu}=\{0,\frac{1}{\sqrt{g_{rr}}},0,0\}
\label{eq13}
\end{equation}
and 
\begin{equation}
\widetilde{\lambda}_{3}^{\mu}=\{\frac{L}{\sqrt{g}},0,0,\frac{E}{\sqrt{g}}\}
\label{eq14}
\end{equation}
As these vectors are not necessarily propagated along the geodesic, we form their linear combination given by:
\begin{equation}
\pmb{\Lambda}=\sigma_1\pmb{\widetilde{\lambda}_1}+\sigma_3\pmb{\widetilde{\lambda}_3}
\label{eq15}
\end{equation}
$\sigma_1$ and $\sigma_3$  depend on proper time, and they satisfy the condition $\sigma_{1}^2+\sigma_{3}^2=1$. 
After putting the condition that $\pmb\Lambda$ is parallely propagated, we get
\begin{equation}
\dot{\sigma}_1\widetilde{\lambda}_{1}^\nu+\dot{\sigma}_3\widetilde{\lambda}_{3}^\nu=
-\sigma_1\lambda_{0}^\mu\nabla\mu\widetilde{\lambda}_{1}^\nu-\sigma_3\lambda_{0}^\mu\nabla\mu\widetilde{\lambda}_{3}^\nu
\label{eq16}
\end{equation}
This in turn gives 
\begin{equation}
\dot{\sigma}_1 =-\sigma_3\lambda_{0}^\mu g_{\rho\nu}\widetilde{\lambda}_{1}^\rho\nabla_{\mu}\widetilde{\lambda}_{3}^\nu,~
\dot{\sigma}_3=-\sigma_1\lambda_{0}^\mu g_{\rho\nu}\widetilde{\lambda}_{3}^\rho\nabla_{\mu}\widetilde{\lambda}_{1}^\nu\\
=\sigma_1\lambda_{0}^\mu g_{\rho\nu}\widetilde{\lambda}_{1}^\rho\nabla_{\mu}\widetilde{\lambda}_{3}^\nu
\label{eq17}
\end{equation}
With $\sigma_1=\cos\Psi$ and $\sigma_3=\sin\Psi$ we can form two linear combinations of the form:
\begin{eqnarray}
\pmb{\lambda_{1}}=\pmb{\widetilde{\lambda}_1}\cos\Psi+\pmb{\widetilde{\lambda}_3}\sin\Psi\nonumber\\
\pmb{\lambda_{3}}=-\pmb{\widetilde{\lambda}_1}\sin\Psi+\pmb{\widetilde{\lambda}_3}\cos\Psi
\label{eq18}
\end{eqnarray}
such that we get,
\begin{equation}
\dot{\Psi}=\Omega=\lambda_{0}^\mu g_{\rho\nu}\widetilde{\lambda}_{1}^\rho\nabla_{\mu}\widetilde{\lambda}_{3}^\nu=-\lambda_{0}^\mu g_{\rho\nu}\widetilde{\lambda}_{3}^\rho\nabla_{\mu}\widetilde{\lambda}_{1}^\nu
\label{eq19}
\end{equation}
For equatorial circular orbits, $\Omega$ is only a function of $r$, and hence a constant for each orbit. 
For the Kerr metric we have $\dot{\Psi}=\sqrt{\frac{M}{r^3}}$, matching with \cite{Ishii}. These four parallely 
propagated vectors ($\pmb{\lambda_{0}}$,$\pmb{\lambda_{1}}$,$\pmb{\lambda_{2}}$,$\pmb{\lambda_{3}}$) can be used 
as coordinate directions for Fermi normal coordinates. Thus we have solved the problem of finding Fermi normal 
coordinates for equatorial circular orbits in a sufficiently general metric.
The transformation from BL to Fermi normal coordinates(\{$x^\mu$\}) is given by:
\begin{equation}
R_{abcd}=R_{\mu\nu\rho\sigma}\lambda^{\mu}_{a}\lambda^{\nu}_{b}\lambda^{\rho}_{c}\lambda^{\sigma}_{d}
\label{eq20}
\end{equation}
Also, to get rid of $\Psi$ from equations, we define coordinate system defined by:
\begin{eqnarray}
\widetilde{x}^1=x^1\cos\Psi+x^3\sin\Psi\\
\widetilde{x}^3=-x^1\sin\Psi+x^3\cos\Psi
\label{eq21}
\end{eqnarray}
$\widetilde{x}^i$ 's are the coordinates corresponding to $\widetilde{\lambda}_a$ frame. 
The central singularity is on the negative $\widetilde{x}^1$ axis.

\subsection{Hydrodynamic equations}

We assume the rotating star to be made of inviscid fluid and having radius much less than $r$, the distance from source.
 Also, we assume its mass to be much less than $M$ thus its own potential can be described by Newtonian gravity and can 
 be linearly superposed with tidal potential. The fluid that makes up the star follows \cite{Ishii} :
\begin{equation}
\rho\frac{\partial v_{i}}{\partial \tau}+\rho v^{j}\frac{\partial v_{i}}{\partial x^{j}} = -\frac{\partial P}{\partial x^{i}}-\rho\frac{\partial (\phi+\phi_{tidal})}{\partial x^{i}}+\rho\Bigg[v_{j}\bigg(\frac{\partial A_{j}}{\partial x^{i}}-\frac{\partial A_{i}}{\partial x^{j}}\bigg)-\frac{\partial A_{i}}{\partial \tau}\Bigg]
\label{eq22}
\end{equation}
Where $v^i$ denote velocity field of the fluid in Fermi normal coordinates. $A_{i}$ has been defined previously. 
The terms in the right hand side of eq.(\ref{eq22}) involving $A_i$ correspond to gravitomagnetic effects. 
$\phi$ is the potential due to self gravity of the star. 
We assume the corotational velocity field given by:
\begin{equation}
 v^i = \Omega\left(-\{x^3-x_c\sin(\Omega\tau)\},0,\{x^1-x_c\cos(\Omega\tau)\}\right)
 \label{eq23}
\end{equation}
where $\Omega$ is defined in eq.(\ref{eq19}). 
The equation of state of the star is assumed to be $P=\kappa\rho^{1+\frac{1}{n}}$. Then, the hydrostatic equilibrium condition,
along with the equation of continuity leads to the Lane-Emden equation given by
\begin{equation}
\frac{1}{\xi^2}\frac{d}{d\xi}\left(\xi^2\frac{d\theta}{d\xi}\right) + \theta^n=0
\label{LaneEmden}
\end{equation}
where $\xi$ is a dimensionless Lane-Emden coordinate obtained by a scaling of the radial coordinate, and $\rho = \rho_c\theta^n$, with 
$\rho_c$ being the central density. If we denote by $R_{0}$ the radius of the star, and $\xi_1$ is the Lane-Emden
coordinate at its surface, then 
\begin{equation}
R_0=\Bigg[\frac{(n+1)\kappa\rho_{c}^{(1-n)/n}}{4\pi}\Bigg]^{\frac{1}{2}}\xi_1
\label{eq24}
\end{equation}
Putting $P$ and $v^i$ in eq.(\ref{eq22}),
integrating and transforming to $\widetilde{x}^i$, we get, with $x_g = 2x_c$,
\begin{equation}
\frac{\Omega^2}{2}[(\widetilde{x}^1-x_g)^2+(\widetilde{x}^3)^2]=\kappa(n+1)\rho^{\frac{1}{n}}+\phi+\phi_{tidal}+\phi_{mag}+C
\label{eq25}
\end{equation}
with $C$ being an integration constant that is fixed from boundary conditions, and 
$\phi_{mag}$ is the potential due to gravitomagnetic effects and comes from integration of terms containing $A_i$'s. 
Also, we have the Poisson's equation:
\begin{equation}
\Delta\phi=4\pi\rho
\label{eq26}
\end{equation}
Then, \eqref{eq25} and \eqref{eq26} are the two equations that are numerically solved in order to get $\rho$ and $\phi$.  
Since the numerical procedure is quite involved, let us now list a few details of this. 

\subsection{\label{sec:level2}Details of numerical procedure}
Our numerical method closely follows the one developed in \cite{Ishii} :
\begin{enumerate}
	\item We choose units such that $M=1$ (we will momentarily comment on restoration of units). It is also assumed that the radius of the star is much
	less than its radial distance $r$ from the source. 
	\item We assume that the celestial object moving in the background of eq.(\ref{eq1}) does not backreact on the metric. This is of importance
	as the rotating JNW metric is not a vacuum solution of the Einstein's equations and is seeded by a scalar field. 
	\item All numerical computations are done including the gravitomagnetic effect. 
	\item For a given $r$, $\gamma$ and $a$, eq.(\ref{eq9}) and eq.(\ref{eq10}) are numerically solved to get $E$ and $L$. 
	\item We have done our computation assuming $n = 1.0$ and $R_{0} = 0.5M$. $\rho_c$ is assumed suitably (discussed later). 
	For $n=1.0$, it can be shown that $\xi_1 = \pi$. Then eq.(\ref{eq24}) gives $\kappa$.
	\item Using $\widetilde{x}^i = \lambda q^i$, $x_g = \lambda q_g$, $\phi = \lambda^2\phi'$, eqs.(\ref{eq25}) and (\ref{eq26}) 
	can be transformed as:
	\begin{equation}
	\frac{\Omega^2}{2}\left((q^1-q_g)^2+(q^3)^2\right)\lambda^2=\kappa(n+1)\rho^{\frac{1}{n}}
	+\lambda^2\left(\phi'+\phi'_{tidal}+\phi'_{mag}\right)+C
	\label{eq27}
	\end{equation}
	\begin{equation}
	\Delta_q\phi'=4\pi\rho
	\label{eq28}
	\end{equation}
	This scaling is done to ensure that the size of star is within bound.
	\item In our calculation, at first we assumed a density distribution. To solve eq.(\ref{eq28}) from it, we 
used a grid of $101$ $\times$ $101$ $\times$ $101$ points. 
Calculation can be done using symmetry thus reducing the number of points by a factor of two. Although in our case this 
did not cause much difference in terms of the computation time, so we proceeded with cubic volume.
	\item We assumed that the size of star is small enough (than size of cube) such that at boundary the potential is given by 
	$-\int_{cube}\frac{\rho d^3q}{r_q'}$. Where $r_q'$ is the distance of boundary from 
	$q^i = (0,0,0)$. The integral is done over whole cube, because outside the star $\rho = 0$ anyway.
	\item Thus \eqref{eq28} is solved using Dirichlet boundary condition; cyclic reduction method is used to solve 
	the corresponding matrix equation.
	\item $(q_s, 0, 0)$ is assumed to be the point on surface of the star nearest to source. $q_s$ is a negative number. 
	Also, the density of the star is maximum at $(0,0,0)$ and has a value $\rho_c$. These conditions, when put in eq.(\ref{eq27}), gives:
	\begin{equation}
	q_g = -\frac{1}{\Omega^2}\Bigg[\frac{\partial}{\partial q^1}[\phi'+\phi'_{tidal}+\phi'_{mag}]\Bigg]_{(0,0,0)}
	\label{eq29}
	\end{equation}  
	Similarly $C$ and $\lambda$ can be computed. 
	\item Then eq.(\ref{eq27}) gives values of $\rho$ at the grid points. Using this $\rho$, step 5 onwards, everything can be repeated. 
	Thus through iterations we can solve for $\rho$ and $\phi'$.
	\item Beyond the Roche limit, the density contours at $(q_s,0,0)$ begin to break. Thus at the Roche limit:
	\begin{equation}
	\frac{\partial\rho}{\partial q^1}\Biggr\rvert_{(q_s,0,0)} = 0
	\label{eq30}
	\end{equation}
	This is actually the condition for the formation of a cusp.
	\item $\rho$ and $\phi'$ are computed for a series of $\rho_c$ values to determine the critical $\rho_c$ value ($\rho_{crit}$). 
	Stars with $\rho_c<\rho_{crit}$ disrupt due to tidal forces. $\rho_{crit}$ is then used to compute $M_{crit}$, the critical mass below
	which the star will disintegrate. Also, $\rho_{crit}$ is used to calculate the dimensionless quantity 
	$\xi_{crit} = \frac{\Omega^2}{\pi\rho_{crit}}$. Clearly, stars with $\xi < \xi_{crit}$ are stable under tidal disruptions. 
\end{enumerate}

It is to be kept in mind that here we mostly deal with dimensionless quantities. 
In order to restore dimensions, we can always insert appropriate factors of the ADM mass, 
along with $c$ and $G$. In the results that we show, the critical mass of the star ($M_{crit}$) is expressed in units of the ADM mass. 
Upon restoring appropriate dimensions, working in S.I units, we have 
\begin{equation}
M_{crit}^{(d)} \mbox{ (kg)} =M_{crit} \times M^{(d)} \mbox{ (kg)}
\end{equation}
where $M^{(d)}$ is the ADM mass (of the central object) (in {\rm kg}), the superscript denoting that dimensions have been restored. 
Similarly, the radius $r$ that we use is related to the quantity $r^{(d)}$ in S.I units as $r^{(d)} = r\times GM^{(d)}/c^2$. This would
also hold for the results that we derived on the frame dragging effect, in the previous section. 

Before we proceed to the results, a few words about the stability of the equatorial circular orbits is in order. 

\subsection{Stability of circular orbits}

It is important to study the stability of circular orbits for which we will compute the tidal effects. For Kerr backgrounds, this has a long
history and relevant details can be found, for example, in the classic paper by Bardee, Press and Teukolsky \cite{BardeenKerr1}
or in Chandrasekhar's textbook. \cite{Chandra}. Even for
the Kerr example, the computations become tedious for generic values of $\theta$, but is somewhat tractable for equatorial 
orbits with $\theta = \pi/2$. Essentially, for a stationary space-time, one writes down an effective potential, 
\begin{equation}
{\dot r}^2 + V\left(r\right) = 0~,~  V(r) = \frac{1}{g_{rr}}\left(g_{tt}{\dot t}^2 + g_{\phi\phi}{\dot \phi}^2 + 2g_{t\phi}{\dot t}{\dot \phi} + 1\right)
\label{veff}
\end{equation}
In \cite{BardeenKerr1}, the effective potential on the equatorial plane is more conveniently multiplied by a factor of $r^4$ compared
to the expression in eq.(\ref{veff}). Then, the conditions for stable circular orbits boil down to 
\begin{equation}
V\left(r\right)=0,~V'\left(r\right)=0,~V''\left(r\right)\geq 0
\label{stabcon}
\end{equation}
where a prime denotes a derivative with respect to the radial coordinate. The first two equations can be solved to obtain the
conserved energy and angular momentum (per unit masses) via eq.(\ref{eq8}), and this is then input in the third to check
for stability. 

First a few words on the known solutions for the non-rotating JNW space-time, obtained from eq.(\ref{eq1}) by setting $a=0$
and $M=1$. In that case, one obtains
\begin{equation}
V\left(r\right) = \frac{\gamma  L^2 \left(1-\frac{2}{\gamma  r}\right)^{2 \gamma }+E^2 r (2
   -\gamma  r)+\left(\gamma  r^2-2  r\right) \left(1-\frac{2 }{\gamma 
   r}\right)^{\gamma }}{r (\gamma  r-2 )}
\end{equation}
It is then not difficult to solve for $E^2$ and $L^2$ from the equations in eq.(\ref{stabcon}). Doing this, one finds (for $M=1$)
that marginally stable orbits for which $V''(r)=0$ occurs for two values of the radial coordinate,
\begin{equation}
r_{\pm} = \frac{1}{\gamma}\left(1+3\gamma \pm \sqrt{5\gamma^2-1}\right)
\end{equation}
Then, it is readily seen that stable circular orbits exist at all radii (greater than $2M=2$) for $\gamma < 1/\sqrt{5}$. A more detailed
analysis for other values of $\gamma$ was done in \cite{AnirbanJoshi} to which we refer the reader for more details. 

In the present case, the rotation parameter complicates the issue, essentially due to the appearance of cross terms involving
$E$ and $L$. With non zero $a$, we have the effective potential
(again with $M=1$) given by
\begin{equation}
V\left(r\right) = \frac{{\mathcal A}^{\gamma } \left(a \gamma  \left(-2 a
   E^2+a+2 EL \right)+r (\gamma  r-2)\right)+\gamma  {\mathcal A}^{2 \gamma } (L-a E)^2-E^2 r (\gamma  r-2)}{r (\gamma  r-2)}
\label{JMNRv}
\end{equation}
where we have defined ${\mathcal A}=\left(1-\frac{2}{\gamma  r}\right)$. The $\gamma = 1$ (Kerr) case is well known, but 
it is difficult to envisage an analytic treatment of eq.(\ref{stabcon}) via eq.(\ref{JMNRv}) for generic $a$ and $\gamma$. 

Although we were unable to compute any analytic expression (even in the Kerr limit, by doing 
a series expansion as $\gamma \to 1$), we have checked in specific cases that we consider here 
that in the regions of interest that we describe below, 
stable circular orbits are possible.\footnote{We remind the reader that a comprehensive analysis of the stability conditions
for circular orbits in the JNW background was performed in \cite{Harko} where it was shown that for moderate to high values 
of $\gamma$, stable circular orbits exist
at all points outside the singular surface, and that marginally stable orbits (innermost stable particle orbits) are allowed for
low values of $\gamma$ only for very slowly rotating JNW singularities.}
This is done by the following method. We choose a specific value of $\gamma$ to solve the first two equations of eq.(\ref{stabcon})
using the effective potential of eq.(\ref{JMNRv}) (or equivalently, use point (4) of subsection 4.3). 
This gives a set of solutions for $E$ and $L$, from which we pick the one for 
which the energy approaches unity in the limit $r\to \infty$ and $L$ is positive, and then test for $V''(r)$. 

\subsection{Results and analysis}

\begin{figure}[h!]
 \centering
 \subfigure[]{
 \includegraphics[width=2.5in,height=2.3in]{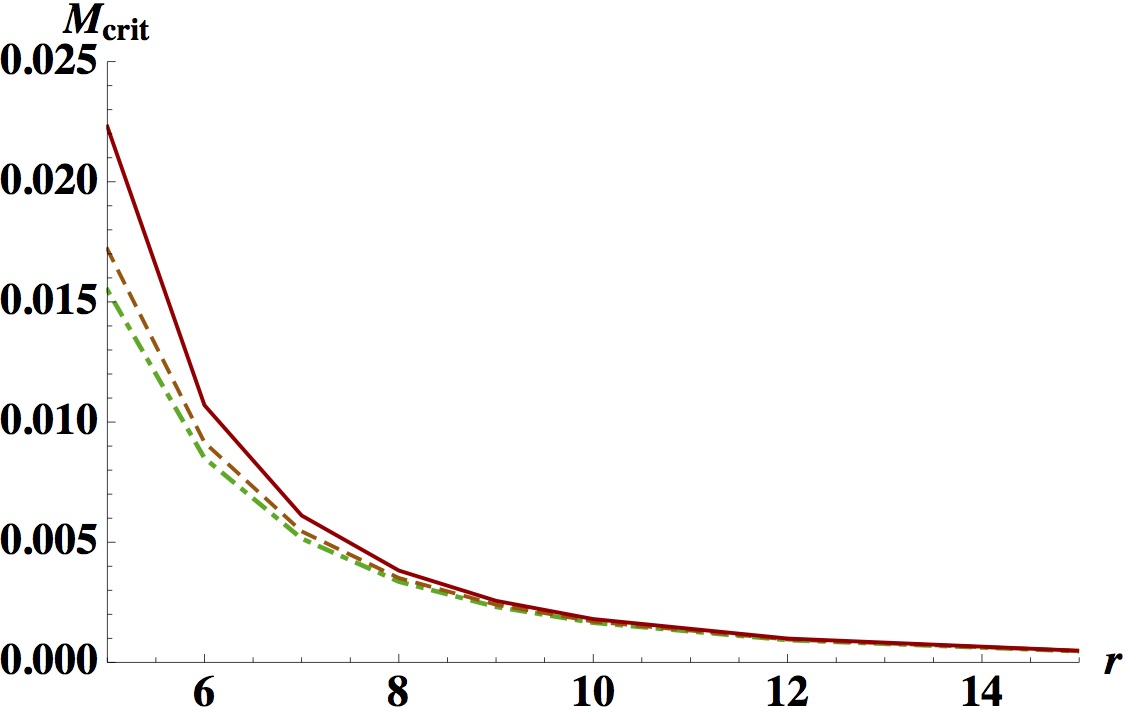}
 \label{tidal1} } 
 \subfigure[]{
 \includegraphics[width=2.5in,height=2.3in]{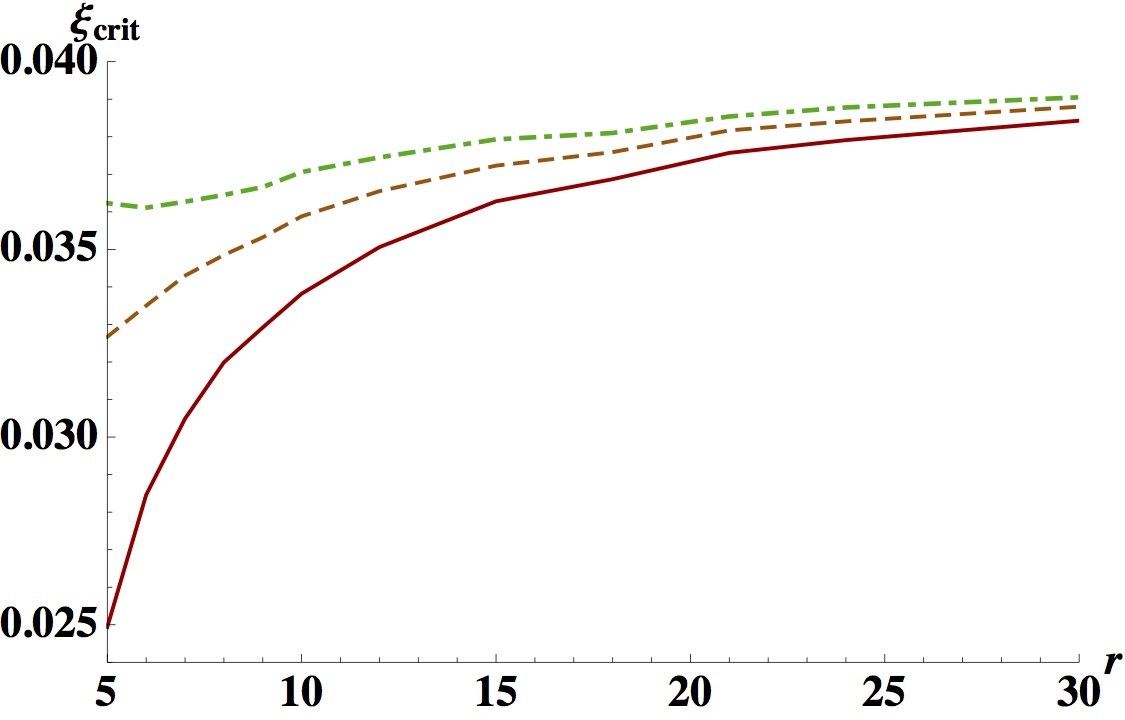}
 \label{tidal2} }
\caption{\small{(a)\, $M_{crit}$ vs $r$ for $\gamma = 0.8$ with $M=1$,for equatorial circular orbits. 
The solid red, dashed orange and dot-dashed
green curves correspond to $a=0.05$, $0.7$ and $1.2$ respectively.  
(b)\, $\xi_{crit}$ vs $r$ for $\gamma = 0.8$ with $M=1$ for equatorial circular orbits. The solid red, dashed orange and dot-dashed
green curves correspond to $a=0.05$, $0.7$ and $1.2$ respectively.}}
 \end{figure}

\begin{figure}[h!]
 \centering
 \subfigure[]{
 \includegraphics[width=2.5in,height=2.3in]{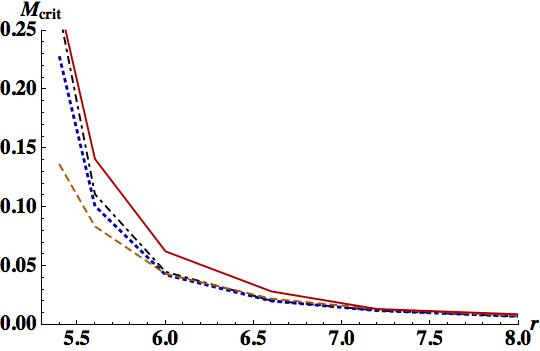}
 \label{tidal3} } 
 \subfigure[]{
 \includegraphics[width=2.5in,height=2.3in]{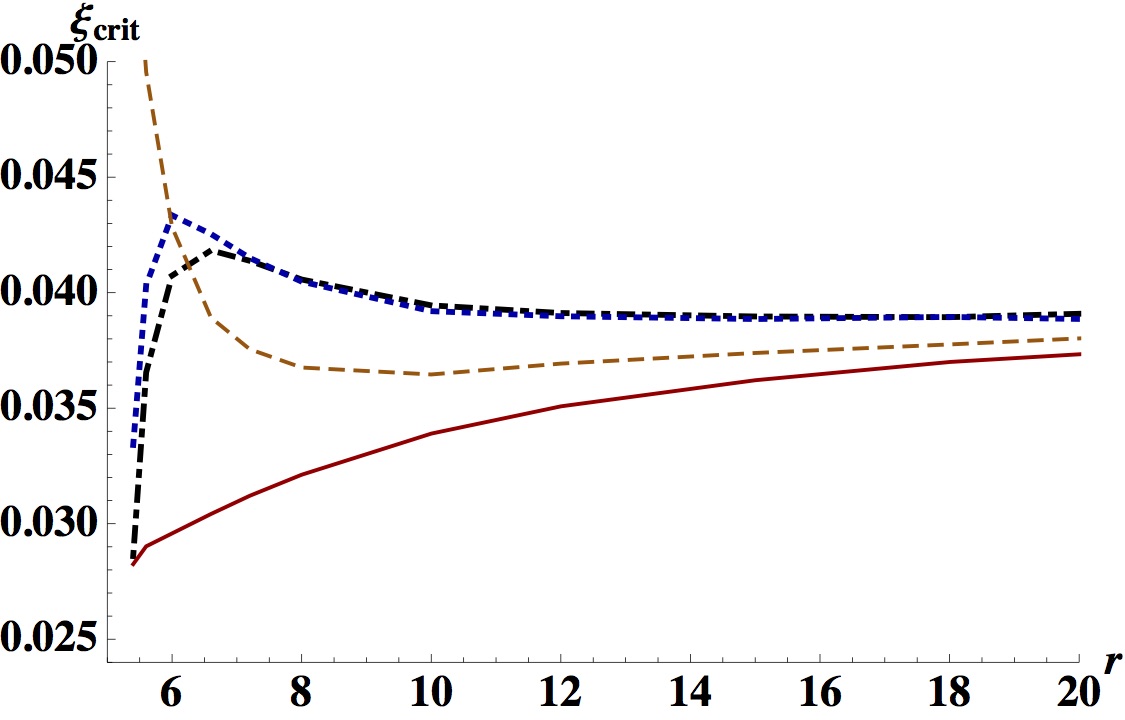}
 \label{tidal4} }
\caption{\small{(a)\, $M_{crit}$ vs $r$ for $\gamma = 0.4$ with $M=1$, for equatorial circular orbits. The solid red, dashed orange, dotted
blue and dot-dashed black curves correspond to $a=0.05$, $0.7$ and $1.8$ and $2$, respectively.  
(b)\, $\xi_{crit}$ vs $r$ for $\gamma = 0.4$ with $M=1$, for equatorial circular orbits. The solid red, dashed orange, dotted blue 
and dot-dashed black curves correspond to $a=0.05$, $0.7$ and $1.8$ and $2$, respectively.}}
 \end{figure}

In this subsection we present our results and analysis of the tidal forces in the background of eq.(\ref{eq1}). Our main findings
are depicted in figs.(\ref{tidal1}), (\ref{tidal2}) for a high value of $\gamma = 0.8$ and figs. (\ref{tidal3}) and (\ref{tidal4}) for
a lower value of $\gamma=0.4$.
In fig.(\ref{tidal1}), we plot numerical results for $M_{crit}$ as a function of the star's radial distance from the origin. 
Here, we have taken $\gamma = 0.8$, so that the naked singularity is now located at $r=2.5$ (in units of $M$), from 
eq.(\ref{NSlocation}). The maximum value of the angular momentum parameter is $a=1.25$. Here, the solid red, 
dashed orange and dot-dashed green curves correspond to $a=0.05$, $0.7$ and $1.2$ respectively. We note that with increase
of angular momentum of the source, $M_{crit}$ steadily decreases, with the effect being more pronounced 
as one goes closer to the location of the singularity. Since stars with mass lower than $M_{crit}$ are tidally disrupted, this
shows that increase in the spin of the naked singularity stabilises the star. Fig.(\ref{tidal2}) shows the corresponding feature
in $\xi_{crit}$. A similar result was obtained in \cite{Ishii} for the Kerr black hole (see fig.(4) of that paper). 

We however find that as one moves deeper into the naked singularity regime (with a smaller value of $\gamma$), the situation
changes. In figs.(\ref{tidal3}) and (\ref{tidal4}), we plot $M_{crit}$ and $\xi_{crit}$ respectively, for $\gamma = 0.4$. In this case, the horizon 
is located at $r=5$ in units of $M$, and one has the maximum value of the angular momentum parameter $a=2.5$. In the figures,
the solid red, dashed orange, dotted blue and dot-dashed black curves correspond to $a=0.05$, $0.7$ and $1.8$ and $2$, respectively.
Fig.(\ref{tidal3}) shows that $M_{crit}$ decreases sharply as the spin of the source is increased from a small value, and then increases
as the spin is further increased. The corresponding situation for $\xi_{crit}$ is depicted in fig.(\ref{tidal4}). 
$\xi_{crit}$ increases with increase in $a$ up to a radius close to the singularity, but beyond this is pulled down to approach smaller
values close to ones for small $a$. Note that indications of this behaviour was already there in \cite{Ishii} (see fig.(7) of that paper),
where $\xi_{crit}$ showed a sharp increase as one approached the outer event horizon of the Kerr black hole, but what we see over
here is that close to the singularity this trend is reversed. Whether a higher order correction in terms of the Fermi normal coordinates
will change the situation is at present not clear to us, and this requires further study. However, within the fourth order framework that
we have worked with here, it is evident that there is a dramatic difference in the nature of tidal forces close to a naked singularity, compared
to the black hole case. 

To substantiate the discussion above, we will now provide comparitive tables for $M_{crit}$, $\xi_{crit}$ and $\rho_{crit}$ 
for the Kerr and the rotating JNW solutions, with different values of the parameters $a$ and $\gamma$. 
In all these cases, we have chosen the numerical value of $M=1$ and the values of $\xi_{crit}$ have been scaled up by a factor
of $100$ to offer ease of comparison. 
\begin{table}[h!]
\centering
\makebox[0pt][c]{\parbox{1.0\textwidth}{%
\begin{minipage}[b]{0.5\hsize}\centering
 \begin{tabular}{|p{0.3cm}|p{1.8cm}|p{1.8cm}|p{1.5cm}|}
\multicolumn{4}{c}
{}\\
\hline
$r$ & \hspace{0.1cm} $M_{crit}$ & $\xi_{crit} \times 10^2$ & $\rho_{crit}$  \\
\hline
$5$ &\hspace{0.1cm}0.01722&\hspace{0.1cm}2.564 & 0.0993 \\
\hline
$8$ &\hspace{0.1cm}0.00338&\hspace{0.1cm}3.205 & 0.0194 \\
\hline
$12$ &\hspace{0.1cm}0.00092&\hspace{0.1cm}3.515 & 0.0052 \\
\hline
$15$ &\hspace{0.1cm}0.00046&\hspace{0.1cm}3.627 & 0.0026 \\
\hline
$18$ &\hspace{0.1cm}0.00026&\hspace{0.1cm}3.688 & 0.0015 \\
\hline
$21$ &\hspace{0.1cm}0.00016&\hspace{0.1cm}3.756 & 0.0009 \\
\hline
\end{tabular}
\caption{$M_{crit}$, $\xi_{crit}$ and $\rho_{crit}$, for the Kerr black hole with $a = 0.05$.} 
\label{table1}
\end{minipage}
\hfill
\begin{minipage}[b]{0.5\hsize}\centering
 \begin{tabular}{|p{0.3cm}|p{1.8cm}|p{1.8cm}|p{1.5cm}|}
\multicolumn{4}{c}
{}\\
\hline
$r$ & \hspace{0.1cm} $M_{crit}$ & $\xi_{crit} \times 10^2$ & $\rho_{crit}$  \\
\hline
$5$ &\hspace{0.1cm}0.01318&\hspace{0.1cm}3.368 & 0.0756 \\
\hline
$8$ &\hspace{0.1cm}0.00306&\hspace{0.1cm}3.552 & 0.0175 \\
\hline
$12$ &\hspace{0.1cm}0.00087&\hspace{0.1cm}3.691 & 0.0050 \\
\hline
$15$ &\hspace{0.1cm}0.00044&\hspace{0.1cm}3.743 & 0.0025 \\
\hline
$18$ &\hspace{0.1cm}0.00025&\hspace{0.1cm}3.790 & 0.0014 \\
\hline
$21$ &\hspace{0.1cm}0.00016&\hspace{0.1cm}3.832 & 0.0009 \\
\hline
\end{tabular}
\caption{$M_{crit}$, $\xi_{crit}$ and $\rho_{crit}$, for the Kerr black hole with $a = 0.9$.} 
\label{table2}
\end{minipage}
}}
\end{table}
\begin{table}[h!]
\centering
\makebox[0pt][c]{\parbox{1.0\textwidth}{%
\begin{minipage}[b]{0.5\hsize}\centering
 \begin{tabular}{|p{0.3cm}|p{1.8cm}|p{1.8cm}|p{1.5cm}|}
\multicolumn{4}{c}
{}\\
\hline
$r$ & \hspace{0.1cm} $M_{crit}$ & $\xi_{crit} \times 10^2$ & $\rho_{crit}$  \\
\hline
$5$ &\hspace{0.1cm}0.0389&\hspace{0.1cm}2.459 & 0.2250 \\
\hline
$8$ &\hspace{0.1cm}0.0048&\hspace{0.1cm}3.194 & 0.0276 \\
\hline
$12$ &\hspace{0.1cm}0.0011&\hspace{0.1cm}3.509 & 0.0064 \\
\hline
$15$ &\hspace{0.1cm}0.00053&\hspace{0.1cm}3.627 & 0.0030 \\
\hline
$18$ &\hspace{0.1cm}0.00029&\hspace{0.1cm}3.708 & 0.0017 \\
\hline
$21$ &\hspace{0.1cm}0.00018&\hspace{0.1cm}3.784 & 0.0010 \\
\hline
\end{tabular}
\caption{$M_{crit}$, $\xi_{crit}$ and $\rho_{crit}$, for rotating JNW, with $\gamma = 0.6$, $a = 0.05$.} 
\label{table3}
\end{minipage}
\hfill
\begin{minipage}[b]{0.5\hsize}\centering
 \begin{tabular}{|p{0.3cm}|p{1.8cm}|p{1.8cm}|p{1.5cm}|}
\multicolumn{4}{c}
{}\\
\hline
$r$ & \hspace{0.1cm} $M_{crit}$ & $\xi_{crit} \times 10^2$ & $\rho_{crit}$  \\
\hline
$5$ &\hspace{0.1cm}0.0258&\hspace{0.1cm}3.738 & 0.1480 \\
\hline
$8$ &\hspace{0.1cm}0.00427&\hspace{0.1cm}3.613 & 0.0244 \\
\hline
$12$ &\hspace{0.1cm}0.00107&\hspace{0.1cm}3.704 & 0.0061 \\
\hline
$15$ &\hspace{0.1cm}0.00515&\hspace{0.1cm}3.763 & 0.0029 \\
\hline
$18$ &\hspace{0.1cm}0.00029&\hspace{0.1cm}3.823 & 0.0016 \\
\hline
$21$ &\hspace{0.1cm}0.00018&\hspace{0.1cm}3.837 & 0.0010 \\
\hline
\end{tabular}
\caption{$M_{crit}$, $\xi_{crit}$ and $\rho_{crit}$, for rotating JNW with $\gamma = 0.6$, $a = 0.9$.} 
\label{table4}
\end{minipage}
}}
\end{table}
In tables (\ref{table1}) and (\ref{table2}), we present the results for the Kerr black hole , with $a=0.05$ and $a=0.9$, respectively.\footnote{The
first entry in table (\ref{table1}) is actually an unstable equatorial circular orbit, as follows from the known equation for the innermost stable circular orbit
for the Kerr space-time, given by (see, e.g \cite{BardeenKerr1}) $r^2 - 6Mr + 8aM^{1/2}r^{1/2} - 3a^2 = 0$, which translates to $r= 5.84$ with $a=0.05$
and $M=1$. Nevertheless, we keep this for comparison purpose. For $a=0.9$, the innermost stable circular orbit occurs at $r=2.32$.}
This is done for similar values of the radial coordinate. In tables (\ref{table3}) and (\ref{table4}), 
the analysis is repeated, again for similar values of $r$ for the rotating JNW space-time with $\gamma = 0.6$, again with $a=0.05$ and $0.9$.
In this case, the naked singularity occurs at $r=3.33$, from eq.(\ref{NSlocation}). 

Comparing the values of tables (\ref{table1}) -- (\ref{table4}), the effect of $\gamma$, the JNW parameter is clearly seen to enhance $M_{crit}$ and 
$\rho_{crit}$ near the location of the singularity, as compared to the Kerr example. For large values of the radial coordinate, this difference ceases
to exist, as expected. 

More drastic differences emerge as we further lower the value of $\gamma$, i.e go deeper into the naked singularity regime. In order to explain
the plots of figs.(\ref{tidal3}) and (\ref{tidal4}), we now present numerical values of $M_{crit}$, $\xi_{crit}$ and $\rho_{crit}$ for $\gamma = 0.4$. In this
case, the location of the singularity is at $r=5$, and the rotation parameter can range from zero to $a=2.5$. 
To compare with figs.(\ref{tidal3}) and (\ref{tidal4}),
we tabulate the data for the physical parameters for $a=0.05$, $0.7$, $1.8$ and $2.0$, in tables (\ref{table5}), (\ref{table6}), (\ref{table7}) and
(\ref{table8}), respectively. 
\begin{table}[h!]
\centering
\makebox[0pt][c]{\parbox{1.0\textwidth}{%
\begin{minipage}[b]{0.5\hsize}\centering
 \begin{tabular}{|p{0.4cm}|p{1.8cm}|p{1.8cm}|p{1.5cm}|}
\multicolumn{4}{c}
{}\\
\hline
$r$ & \hspace{0.1cm} $M_{crit}$ & $\xi_{crit} \times 10^2$ & $\rho_{crit}$  \\
\hline
$5.4$ &\hspace{0.1cm}0.272&\hspace{0.1cm}2.825 & 1.620 \\
\hline
$5.6$ &\hspace{0.1cm}0.141&\hspace{0.1cm}2.902 & 0.835 \\
\hline
$6.0$ &\hspace{0.1cm}0.063&\hspace{0.1cm}2.957 & 0.365 \\
\hline
$6.6$ &\hspace{0.1cm}0.029&\hspace{0.1cm}3.042 & 0.165 \\
\hline
\end{tabular}
\caption{$\gamma = 0.4,~a = 0.05$.}
\label{table5}
\end{minipage}
\hfill
\begin{minipage}[b]{0.5\hsize}\centering
 \begin{tabular}{|p{0.4cm}|p{1.8cm}|p{1.8cm}|p{1.5cm}|}
\multicolumn{4}{c}
{}\\
\hline
$r$ & \hspace{0.1cm} $M_{crit}$ & $\xi_{crit} \times 10^2$ & $\rho_{crit}$  \\
\hline
$5.4$ &\hspace{0.1cm}0.136&\hspace{0.1cm}5.728 & 0.798 \\
\hline
$5.6$ &\hspace{0.1cm}0.084&\hspace{0.1cm}4.955 & 0.489 \\
\hline
$6.0$ &\hspace{0.1cm}0.043&\hspace{0.1cm}4.284 & 0.252 \\
\hline
$6.6$ &\hspace{0.1cm}0.022&\hspace{0.1cm}3.892 & 0.129 \\
\hline
\end{tabular}
\caption{$\gamma = 0.4,~a = 0.7$.}
\label{table6}
\end{minipage}
}}
\end{table}
\begin{table}[h!]
\centering
\makebox[0pt][c]{\parbox{1.0\textwidth}{%
\begin{minipage}[b]{0.5\hsize}\centering
 \begin{tabular}{|p{0.4cm}|p{1.8cm}|p{1.8cm}|p{1.5cm}|}
\multicolumn{4}{c}
{}\\
\hline
$r$ & \hspace{0.1cm} $M_{crit}$ & $\xi_{crit} \times 10^2$ & $\rho_{crit}$  \\
\hline
$5.4$ &\hspace{0.1cm}0.227&\hspace{0.1cm}3.341 & 1.370 \\
\hline
$5.6$ &\hspace{0.1cm}0.101&\hspace{0.1cm}4.032 & 0.601 \\
\hline
$6.0$ &\hspace{0.1cm}0.043&\hspace{0.1cm}4.335 & 0.249 \\
\hline
$6.6$ &\hspace{0.1cm}0.020&\hspace{0.1cm}4.254 & 0.118 \\
\hline
\end{tabular}
\caption{$\gamma = 0.4,~a = 1.8$}
\label{table7}
\end{minipage}
\hfill
\begin{minipage}[b]{0.5\hsize}\centering
 \begin{tabular}{|p{0.4cm}|p{1.8cm}|p{1.8cm}|p{1.5cm}|}
\multicolumn{4}{c}
{}\\
\hline
$r$ & \hspace{0.1cm} $M_{crit}$ & $\xi_{crit} \times 10^2$ & $\rho_{crit}$  \\
\hline
$5.4$ &\hspace{0.1cm}0.264&\hspace{0.1cm}2.860 & 1.600 \\
\hline
$5.6$ &\hspace{0.1cm}0.111&\hspace{0.1cm}3.655 & 0.663 \\
\hline
$6.0$ &\hspace{0.1cm}0.045&\hspace{0.1cm}4.074 & 0.265 \\
\hline
$6.6$ &\hspace{0.1cm}0.002&\hspace{0.1cm}4.183 & 0.120 \\
\hline
\end{tabular}
\caption{$\gamma = 0.4,~a = 2.0$}
\label{table8}
\end{minipage}
}}
\end{table}
These numbers are readily compared with those in tables (\ref{table1}) -- (\ref{table4}), and it is seen that the effect of the naked singularity
can enhance $M_{crit}$ and $\rho_{crit}$ by more than an order of magnitude, close to the singularity. 
In these tables, we have restricted attention to stable circular orbits near the singularity. We note here that if the mass of the central 
singularity is high, then this enhancement of the critical mass might assume some observational significance in future, but this is at best
a speculation.

\section{Discussions and conclusions}

Distinguishing black holes from naked singularities has been a topic of much discussion over the last few years. Although singularities
in general indicate the limits of applicability of a theory, the discussion assumes relevance in the context of quantum effects setting
in for strong gravity. This is also particularly interesting as over the years evidence has emerged that naked singularities can indeed
form out of a gravitational collapse process, also in a cosmological setting. 

In this paper, we have taken a few moderate steps in this study. Our focus was on two directions : the Lense-Thirring precession and
tidal effects, both in the strong gravity regime. In the former, we saw a few features that might be indicative of interesting physics :
firstly, contrary to the weak field L-T precession, we saw that there is a possibility of the L-T precession frequency increasing (for a Copernican 
static observer just outside the ergoregion of a Kerr black hole) as one decreases
the angular momentum of the black hole. This feature persisted for the rotating JNW background. 
We established that for the black hole, such an increase is possible in a physical process like the Penrose process, where the black hole
slows down as energy is extracted from it. This feature was notably absent for the accreting Kerr black hole. 

We then saw that for generic stationary Copernican observers in the Kerr background, there are certain values of its angular velocity for 
which the precession frequency of a test gyroscope vanishes exactly. 
This feature was present for the accreting Kerr black hole as well. This would in particular mean that Copernican
observers at certain radii outside the black hole should see that the L-T precession frequency in their frame change sign,
as the black hole angular momentum (and mass) changes. 
For Copernican static observers in the rotating JNW background, similar features as the static Kerr observer 
were obtained. In this case, we showed that
the L-T precession frequency might be substantially higher than those of black holes, maybe by almost an order of magnitude. 

We then went on to study a further aspect of great importance in GR -- tidal forces. Following \cite{Ishii}, this was done in Fermi normal
coordinates up to a fourth order expansion in such coordinates. After providing a relatively simple method of constructing Fermi normal
coordinates for circular geodesics in the equatorial plane (this is the only case which can be handled anyway), we discussed tidal effects
in rotating JNW backgrounds. Our numerical results here show a drastic difference with the corresponding Kerr case near the singular 
surface up to the order of approximation that we have assumed. 

To summarize, in this paper we have taken some modest steps that we hope will add to the literature on the distinction between 
black holes and naked singularities. In addition, we have given some interesting aspects of gyroscope precession in Kerr backgrounds. 
Of importance will be to analyze other rotating naked singularity backgrounds to see if these 
results persist. It should also be interesting to go beyond the fourth order approximation of \cite{Ishii} for computing tidal forces. 
Of course this is a substantially difficult task, and is left for a future publication.

\end{document}